\documentclass[11pt,reqno]{article}
\usepackage{bm}
\usepackage{amsthm}
\usepackage{amsmath}
\usepackage{amssymb}
\usepackage[round]{natbib}
\usepackage{color}
\usepackage{authblk}
\usepackage{microtype}
\usepackage{graphicx}
\usepackage{float}
\usepackage{prodint}
\usepackage{paralist}
\usepackage{tabularx}
\usepackage{epstopdf}
\usepackage{rotating}
\usepackage{geometry}
\DeclareMathSymbol{\shortminus}{\mathbin}{AMSa}{"39}
\newcommand{\rnc}{\renewcommand}
\newcommand{\nc}{\newcommand}

\nc{\mb}{\mathbb}
\nc{\mc}{\mathcal}
\nc{\N}{\mb{N}}
\nc{\R}{\mb{R}}
\nc{\Q}{\mb{Q}}

\nc{\E}{E}
\rnc{\P}{P}
\nc{\var}{ \text{Var} }
\nc{\mbf}{\boldsymbol}
\nc{\I}{I}

\nc{\trans}{^{\top}}
\nc{\assumption}{{Assumption A}}

\newtheorem{lemma}{Lemma}
\newtheorem{assump}{Assumption}
\newtheorem{theorem}{Theorem}

\newtheorem{definition}{Definition}

\newcommand\blfootnote[1]{%
  \begingroup
  \renewcommand\thefootnote{}\footnote{#1}%
  \addtocounter{footnote}{-1}%
  \endgroup
} 

\begin{document}

\title{\Large \bf
Permutation test for the multivariate coefficient of variation in factorial designs}
\author[1]{Marc Ditzhaus}
\author[2,$^*$]{\L ukasz Smaga}

\affil[1]{Faculty of Statistics, TU Dortmund University, Germany.}
\affil[2]{Faculty of Mathematics and Computer Science, Adam Mickiewicz University, Poland}

\maketitle

\begin{abstract}
\blfootnote{${}^*$ e-mail: {ls@amu.edu.pl}}
New inference methods for the multivariate coefficient of variation and its reciprocal, the standardized mean, are presented. While there are various testing procedures for both parameters in the univariate case, it is less known how to do inference in the multivariate setting appropriately. There are some existing procedures but they rely on restrictive assumptions on the underlying distributions. We tackle this problem by applying Wald-type statistics in the context of general, potentially heteroscedastic factorial designs. In addition to the $k$-sample case, higher-way layouts can be incorporated into this framework allowing the discussion of main and interaction effects. The resulting procedures are shown to be asymptotically valid under the null hypothesis and consistent under general alternatives. To improve the finite sample performance, we suggest permutation versions of the tests and shown that the tests' asymptotic properties can be transferred to them. An exhaustive simulation study compares the new tests, their permutation counterparts and existing methods. To further analyse the differences between the tests, we conduct two illustrative real data examples.
\end{abstract}



\section{Introduction}
\label{sec:intro}

A widely used unit-free measure of dispersion is the coefficient of variation (CV), which is the ratio of the standard deviation and the population mean. It is a popular tool to judge, e.g., the repeatability of measurements in clinical trials \citep{FeltzMiller1996}, the risk in the financial world \citep{Ferri1979} or in psychology \citep{Weber2004}, and the quantitative variability in genetics \citep{Wright1952}. Moreover, it serves as a reliability tool in control charts \citep{Castagliola:2013,abbasi2018multivariate,nguyen2019one}. The reciprocal of the CV, the standardized mean, is a quantity of its own interest, which can be motivated from one-way analysis of variance problem when the observations are standardized with the sample standard deviation before statistical analysis.

Various inference methods are suggested to compare two or several groups in terms of CV, or equivalently of standardized means. To get an overview, we refer to \cite{AertsHaesbroeck2017} and \cite{PaulySmaga2020}. In various fields, e.g. in biomedicine or psychology \citep{gissi:1990,baigent:etal:1998,cassidy:etal:2008,mehta:etal:2010,kurz:etal:2015}, the one-way layout is too narrow and factorial designs are needed to discuss main effects of different factors, e.g. gender, measurement, site, but also interaction effects between them:  '\textit{it is desirable for reports of factorial trials to include estimates of the interaction between the treatments}' \citep{lubsen:pocock:1994}. Consequently, the question arises: can we extend the existing methods to general factorial designs?

But first, let us come to the multivariate setting. When more than one feature is of interest, comparisons based on marginal CVs are misleading due to potentially different decision for the single features, as pointed out by \citep{VanValen1974} in the biology field, and does not account for correlations between the features.
The solution is to use a summarizing measure for all features, e.g. the multivariate coefficient of variation (MCV). However, a drawback in this direction is that the extension is not unique and there is no default choice up until now. For example, \cite{Reyment1960}, \cite{VanValen1974}, \cite{voinovNikulin1996} and \cite{AlbertZhang2010} suggest to define the MCV by
\begin{align}\label{eqn:MCV}
\sqrt{\frac{(\det\mbf{\Sigma})^{1/d}}{\mbf{\mu}^{\top}\mbf{\mu}}},\ \sqrt{\frac{\mathrm{tr}\mbf{\Sigma}}{\mbf{\mu}^{\top}\mbf{\mu}}},\ \sqrt{\frac{1}{\mbf{\mu}^{\top}\mbf{\Sigma}^{-1}\mbf{\mu}}},\ \sqrt{\frac{\mbf{\mu}^{\top}\mbf{\Sigma}\mbf{\mu}}{(\mbf{\mu}^{\top}\mbf{\mu})^2}},
\end{align}
respectively. Here $\mbf{\mu}$ denotes the nonzero mean vector of a $d$-dimensional random variable and $\mbf{\Sigma}$ is corresponding covariance matrix. All these definitions reduces to the CV in the univariate $(d=1)$ case. The differences of them are discussed in great detail by \cite{AlbertZhang2010}. A further problem of the MCV is the lack of generally applicable inference methods. To the best of our knowledge, there is only a proposal by \cite{AertsHaesbroeck2017} for testing the equality of several MCVs following the definition of \cite{voinovNikulin1996}. But their methods rely on the specific assumption of the underlying distribution and  the convergence speed of their test statistic is rather slow leading to an unstable type-1 error control for small sample sizes; the latter is demonstrated in our simulation study. 

To address all problems raised in the last two paragraphs simultaneously, we suggest Wald-type statistics leading to generally applicable testing procedures 
\begin{enumerate}[(i)]
	\item not relying on any specific distribution assumption. 
	
	\item within the general framework of factorial designs allowing the discussion of main and interaction effects.
	
	\item based on the MCV of \cite{voinovNikulin1996} for treating univariate as well as multivariate settings.
	
	\item being theoretically valid while possessing an accurate type-1 error control under small sample sizes.
\end{enumerate}
We tackle the last aim by following a permutation strategy. It is well-known that permuting exchangeable data (e.g. the distributions in all groups coincide) leads to finitely exact tests. But permutation tests can also be applied beyond the too narrow exchangeability assumption. The finite exactness cannot be preserved but permuted studentized statistics were shown to be still asymptotically exact for various non-exchangeable two-sample scenarios  \citep{neuhaus:1993,janssen:1997studentized,janssenPauls2003,pauly:2011:discussion}.  Recently, the success story of this idea has been continued in the framework of one-way layouts \citep{chung2013exact,chung2016multivariate} and even general factorial designs \citep{paulyETAL2015,friedrich2017permuting, smaga2017,harrar2019comparison,ditzhausETAL2019,dobler:pauly:2019}. For the latter, Wald-type statistics, as proposed here, are favorable choices for such appropriately studentized statistics. In the univariate one-way layout, our proposal coincides with the permutation test of \cite{PaulySmaga2020}.

The remainder of this paper is organized as follows. In Section \ref{sec:setup}, we introduce the general factorial design set-up and formulate the statistical hypotheses in terms of MCVs and standardized means. Moreover, consistent estimators of both are presented. These estimators are used to build the Wald-type statistics in Section \ref{sec:Waldtest}, which are shown to be asymptotically exact under the null hypotheses and consistent under general alternatives. Their permutation counterparts are considered in Section \ref{sec:permutation} and the tests' asymptotic properties are transferred to them. An exhaustive simulation study and illustrative real data examples are presented in Sections \ref{sec:simulation} and \ref{sec:dataexample}, respectively. Section \ref{sec:conclusion} concludes the paper and discusses further research possibilities. All proofs are presented in the appendix.

\section{The set-up}
\label{sec:setup}
We consider the general set-up of mutually independent $d$-dimensional random variables
\begin{align*}
\mbf{X}_{ij} = (X_{ij1}, \ldots, X_{ijd})\trans,
\end{align*}
where the observations $\mbf{X}_{ij}$ have the same distribution for each individual $j=1,\ldots,n_i$ within the group $i=1,\ldots,k$. To include general factorial designs in this framework, the group index $i$ is split up accordingly. For example, let us consider, for a moment, a two-way layout with factors $A$ and $B$ having $a$ and $b$ levels, respectively. Then the group index has the form $i=(i_A,i_B)$ for $i_A=1,\ldots,a$ and $i_B=1,\ldots,b$, and the number of groups $k$ equals $a\cdot b$. Higher-way layouts or nested designs can be incorporated similarly; we refer the reader to \cite{paulyETAL2015} for more details.

Throughout, we assume that all fourth moments $\E(X_{ijr}^4)<\infty$ exists. Moreover, we denote by $\mbf{\mu}_i\neq \mbf{0}$ and $\mbf{\Sigma}_i$ the non-zero expectation vector and the regular covariance matrix of $\mbf{X}_{i1}$, respectively. Following \cite{voinovNikulin1996} we study the multivariate extension of the coefficient of variation given by 
\begin{align*}
C_i = 1/\sqrt{\mbf{\mu}_i\trans\mbf{\Sigma}_i^{-1}\mbf{\mu}_i}.
\end{align*}
We want to point out that by under the present assumptions $C_i$ is always well-defined. 
In the same way, we extend the standardized means to the multivariate setting as follows
\begin{align*}
{B}_i = \sqrt{\mbf{\mu}_i\trans \mbf{\Sigma}_i^{-1}\mbf{\mu}_i}
\end{align*}
Both parameters can be naturally estimated by 
\begin{align*}
\widehat C_i = 1/\sqrt{\mbf{ \widehat\mu}_i\trans \mbf{\widehat \Sigma}_i^{-1}\mbf{\widehat\mu}_i}, \quad \widehat {B}_i = \sqrt{\mbf{ \widehat\mu}_i\trans \mbf{\widehat \Sigma}_i^{-1}\mbf{\widehat\mu}_i},
\end{align*}
where $\mbf{\mu}_i$ and $\mbf{\Sigma}_i$ are replaced by their empirical counterparts 
\begin{align*}
\mbf{\widehat\mu}_i = \frac{1}{n_i}\sum_{j=1}^{n_i} \mbf{X}_{ij},\; \mbf{\widehat \Sigma_i} =\frac{1}{n_i}\sum_{j=1}^{n_i} (\mbf{X}_{ij} - \mbf{\widehat \mu}_i)(\mbf{X}_{ij} - \mbf{\widehat \mu}_i)\trans
\end{align*}

For a given contrast matrix $\mbf{H}\in \R^{r \times k}$, we like to infer the general null hypotheses
\begin{align}\label{eqn:null}
\mathcal H_{0,C}: \mbf{H}\mbf{C} = \mbf{0}, \quad \mathcal H_{0,{B}}: \mbf{H}\mbf{{B}} = \mbf{0},
\end{align}
where $\mbf{C}=(C_1,\ldots,C_k)\trans $ and $\mbf{{B}}$, $\mbf{\widehat C}$, $\mbf{\widehat {B}}$ are defined analogously. Here, $\mbf{H}$ is called a contrast matrix if $\mbf{H}\mbf{1} = \mbf{0}$, and $\mbf{0}$ and $\mbf{1}$ are vectors consisting of $0$'s and $1$'s only. The contrast matrix $\mbf{H}$ is chosen according to the concrete testing problem of interest. Choosing $\mbf{H}= \mbf{P}_k$, where $\mbf{P_k} = \mbf{I}_k  - \mbf{J_k}/k$ is the difference of the unity matrix $\mbf{I}_k$ and the scaled version of the matrix $\mbf{J}_k=\mbf{1}\mbf{1}\trans \in R^{k\times k}$ consisting of $1$'s only, leads to the null hypotheses of no group effect, i.e.,
\begin{align*}
\mathcal H_{0,C}: \{ \mbf{P}_k \mbf{C} = \mbf{0} \} = \{C_1= \ldots = C_k\}.
\end{align*}
How to choose $\mbf{H}$ in a two-way layout to test for no main effects or no interaction effects is explained in Section \ref{sec:simu_fac+design}. The extension to higher-way layouts is straightforward and a brief explanation what to do in hierarchical designs with nested factors is given in Section 4 of \cite{paulyETAL2015}.

For the statistical analysis, the projection matrix $\mbf{T} = \mbf{H}\trans  ( \mbf{H} \mbf{H}\trans )^+ \mbf{H}$ is usually preferred over $\mbf{H}$ itself \citep{brunner:dette:munk:1997,smaga2017,ditzhausETAL2019,dobler:pauly:2019}, where $( \mbf{H} \mbf{H}\trans )^+$ denotes the Moore--Penrose inverse of $\mbf{H} \mbf{H}\trans $.
It is easy to check that both matrices describe the same null hypothesis, but $\mbf{T}$ has some favorable properties as being unique, symmetric and idempotent. Nevertheless, all the results hold independently whether $\mbf{T}$ or $\mbf{H}$ is chosen.

\section{Wald-type test statistic}
\label{sec:Waldtest}

For all asymptotic considerations, we suppose that there are non-vanishing groups in terms of their sample size:
\begin{align}\label{eqn:ni/n}
\frac{n_i}{n} \to \kappa_i \in (0,1).
\end{align}
Here and subsequently, all limits are meant as $n=\sum_{i=1}^kn_i$ tends to $\infty$. To obtain appropriate test statistics for \eqref{eqn:null}, we first derive central limit theorems for the estimates $\mbf{C}_i$ and $\mbf{{B}}_i$. Asymptotic normality of $\mbf{C}_i$ was already proven by \cite{aertsETAL2018ellitical} under elliptical symmetry and in the univariate case without specific distribution assumption by \cite{PaulySmaga2020}. To formulate the corresponding results, let us introduce 
\begin{align*}
&\mbf{A}(\mbf{ \mu}_i,\mbf{\Sigma}_i) = \begin{pmatrix} 2 \mbf{\mu}_i\trans \mbf{\Sigma}_i^{-1} - [(\mbf{\mu}_i\trans \mbf{\Sigma}_i^{-1}) \otimes (\mbf{\mu}_i\trans \mbf{\Sigma}_i^{-1})]\mbf{\widetilde D}(\mbf{\mu}_i) \\
-(\mbf{\mu}_i\trans \mbf{\Sigma}_i^{-1}) \otimes (\mbf{\mu}_i\trans \mbf{\Sigma}_i^{-1})  \end{pmatrix}\trans , \\
\end{align*}
where $\otimes$ is the Kronecker product, and the matrices $\mbf{\widetilde D}(\mbf{x})\in \R^{d^2\times d}$ for $\mbf{x}=(x_1,\ldots,x_d)\trans \in \R^d$, $\mbf{\Psi}_{i3}\in \R^{d^2\times d}$ as well as $\mbf{\Psi}_{i4}\in \R^{d^2\times d^2}$, which are given by their entries
\begin{align}\label{eqn:def_tilde_D}
&[\mbf{\widetilde D}(\mbf{x})]_{ad-d+r,s} = - x_r\I\{s=a\neq r\} - 2 x_s \I\{s=r=a\} \nonumber \\
&\phantom{[\mbf{\widetilde D}(\mbf{x})]_{ad-d+r,s} =} - x_a \I\{r=s\neq a\}\\
&[\mbf{\Psi}_{i3}]_{ad-d+r,s} = \E(X_{i1a}X_{i1r}X_{i1s})  ,\nonumber \\
&\phantom{[\mbf{\Psi}_{i3}]_{ad-d+r,s} = } - \E(X_{i1a}X_{i1r})\E(X_{i1s}) \nonumber \\
&[\mbf{\Psi}_{i4}]_{ad-d+r,bd-d+s}  = \E(X_{i1a}X_{i1r}X_{i1b}X_{i1s}) \nonumber \\
&\phantom{[\mbf{\Psi}_{i4}]_{ad-d+r,bd-d+s}  =}- \E(X_{i1a}X_{i1r})\E(X_{i1b}X_{i1s})\nonumber 
\end{align}
for $a,b,r,s\in\{1,\ldots,d\}$.  
\begin{theorem}\label{theo:conv_C+beta}
	\begin{enumerate}[(i)]
		\item The estimator $\widehat C_i$ is asymptotically normal,
		\begin{align*}
		&n^{1/2} \Big( {\widehat C}_i - {C}_i \Big) \overset{ d}{\longrightarrow} Z_{i,C}\sim N(0,\sigma^2_{i,C}),
		\end{align*}
		with asymptotic variance $\sigma^2_{i,C}$ equal to
		\begin{align*}
		&\frac{1}{4\kappa_i}(\mbf{ \mu}_i\trans \mbf{\Sigma}_i^{-1}\mbf{\mu}_i)^{-3} \mbf{A}(\mbf{ \mu}_i,\mbf{\Sigma}_i)
		\begin{pmatrix}
		\mbf{ \Sigma}_i & \mbf{\Psi}_{i3}\trans  \\
		\mbf{\Psi}_{i3} & \mbf{\Psi}_{i4}
		\end{pmatrix}
		\mbf{A}(\mbf{ \mu}_i,\mbf{\Sigma}_i)\trans .
		\end{align*}
		
		\item We have asymptotic normality of ${\widehat{B}}_i$,
		\begin{align*}
		&n^{1/2} \Big( {\widehat {B}}_i - {{B}}_i \Big) \overset{ d}{\longrightarrow} Z_{i,C}\sim N(0,\sigma^2_{i,C}),
		\end{align*}
		with asymptotic variance $\sigma^2_{i,{B}} = (\mbf{ \mu}_i\trans \mbf{\Sigma}_i^{-1}\mbf{\mu}_i)^{2} \sigma^2_{i,C}$.
	\end{enumerate}
\end{theorem}
In principle, the limits in the previous theorem may be degenerated, i.e. $\sigma_{1,C}^2=0$, and simultaneously $\sigma_{i,{B}}^2=0$, is possible. But this case just appears in rather unusual settings of conditional degenerated or two-point distributions:
\begin{definition}
	The $r$th coordinate $Y_{r}$ of the multivariate random variable $\mbf{Y}=(Y_1,\ldots,Y_d)\trans \in \R^d$ is said to be \textit{conditionally two-point distributed}  if the support of its conditional distribution given the remaining components $(Y_{s})_{s=1,\ldots,d;s\neq r}$ consists of two points at most, including the degenerated case. 
\end{definition}
Examples fulfilling the aforementioned definition are $(Y_1,Y_1^2)$ or $(Y_1,Y_2,Y_1+Y_2^2)$ for arbitrarily distributed $Y_1,Y_2$, and, of course, $(Y_1,\ldots,Y_d)$ in case of binomial distributed $Y_j$. It turns out that these examples need to be excluded to guarantee positive variances $\sigma_{i,C}^2>0$ and $\sigma_{i,{B}}^2>0$:
\begin{lemma}\label{lem:var_equal_zero}
	If $\sigma^2_{i,C} = 0$ holds for some group $i$ then at least one component of $\mbf{X}_{i1}$ is conditionally two-point distributed.
\end{lemma}
Hence, to ensure $\sigma^2_{i,C}>0$ for all $i$, we suppose throughout:
\begin{assump}
	For every group $i$, no component of $\mbf{X}_{i1}$ is conditionally two-point distributed.
\end{assump}
With this assumption at hand, we are now ready to formulate the \textit{Wald-type statistics} (WTS) for testing the null hypotheses \eqref{eqn:null} 
\begin{align*}
&S_{n,C}(\mbf{T}) = n (\mbf{T}\mbf{\widehat C})\trans (\mbf{T}\mbf{\widehat \Sigma_C} \mbf{T}\trans )^+\mbf{T}\mbf{\widehat C}, \\
&S_{n,{B}}(\mbf{T}) = n (\mbf{T}\mbf{\widehat {B}})\trans (\mbf{T}\mbf{\widehat \Sigma_{B}} \mbf{T}\trans )^+\mbf{T}\mbf{\widehat {B}},
\end{align*}
where $\mbf{\widehat\Sigma}_C = \text{diag}(\widehat\sigma^2_{1,C},\ldots,\widehat\sigma^2_{k,C})$ as well as $\mbf{\widehat\Sigma}_{B} = \text{diag}(\widehat\sigma^2_{1,{B}},\ldots,\widehat\sigma^2_{k,{B}})$ are diagonal matrices. Here, $\widehat \sigma^2_{i,C}$ and $\widehat \sigma^2_{i,{B}}$ are the natural estimator of $\sigma^2_{i,C}$ and $\sigma^2_{i,{B}}$, respectively,
\begin{align*}
&\widehat\sigma_{i,C}^2 = \frac{n(\mbf{ \widehat\mu}_i\trans \mbf{\widehat\Sigma}_i^{-1}\mbf{\widehat\mu}_i)^{-3}}{4n_i} \mbf{A}(\mbf{\widehat \mu}_i,\mbf{\widehat\Sigma}_i)
\begin{pmatrix}
\mbf{ \widehat\Sigma}_i & \mbf{\widehat\Psi}_{i3}\trans  \\
\mbf{\widehat\Psi}_{i3} & \mbf{\widehat\Psi}_{i4}
\end{pmatrix}
\mbf{A}(\mbf{ \widehat\mu}_i,\mbf{\widehat\Sigma}_i)\trans , \\
&\widehat\sigma_{i,{B}}^2 = (\mbf{ \widehat\mu}_i\trans \mbf{\widehat\Sigma}_i^{-1}\mbf{\widehat\mu}_i)^{2}\widehat\sigma_{i,C}^2,
\end{align*}
obtained by replacing all expectations by their empirical counterparts, e.g. the entries of $\mbf{\Psi}_{i3}$ by
\begin{align*}
[\mbf{\widehat\Psi}_{i3}]_{ad-d+r,s} = & \Big(n_i^{-1}\sum_{j=1}^{n_i}X_{ija}X_{ijr}X_{ijs}\Big) \\
&- \Big(n_i^{-1}\sum_{j=1}^{n_i}X_{ija}X_{ijr}\Big)\Big(n_i^{-1}\sum_{j=1}^{n_i}X_{ijs} \Big).
\end{align*}
It is straightforward to see that $\mbf{\widehat \Sigma}_C$ and $\mbf{\widehat \Sigma}_{B}$ are consistent estimators of $\mbf{\Sigma}_C=\text{diag}(\sigma_{1,C}^2,\ldots,\sigma_{k,C}^2)$ and $\mbf{\Sigma}_{B} = \text{diag}(\sigma_{1,{B}}^2,\ldots,\sigma_{k,{B}}^2)$. Consequently, Theorem \ref{theo:conv_C+beta} implies convergence in distribution of $S_{n,C}(\mbf{T})$ to  $S_C=\mbf{Z}_C\trans \mbf{T}\trans (\mbf{T}\mbf{\Sigma}_C\mbf{T}\trans )^+\mbf{T}\mbf{Z}_C$ under $\mathcal H_{0,C}: \mbf{T}\mbf{C} = \mbf{0}$, where $\mbf{Z}_C=(Z_{1,C},\ldots,Z_{k,C})\trans $. By Theorem 9.2.2 of \cite{raoMitra1971}, the limit $S_C$ is chi-square distributed with $\text{rank}(\mbf{T})$ degrees of freedom. The same argumentation can be used to derive the limit of $S_{n,{B}}(\mbf{T})$. Both can be summarized as
\begin{theorem}\label{theo:WTS_null}
	Suppose \assumption.
	\begin{enumerate}[(i)]
		\item Under $\mathcal H_{0,C}: \mbf{T}\mbf{C} = \mbf{0}$, $S_{n,C}(\mbf{T})$ converges in distribution to $Z\sim \chi^2_{\text{rank}(\mbf{T})}$.
		
		\item Under $\mathcal H_{0,{B}}: \mbf{T}\mbf{{B}} = \mbf{0}$, $S_{n,{B}}(\mbf{T})$ converges in distribution to $Z\sim \chi^2_{\text{rank}(\mbf{T})}$.
	\end{enumerate}
\end{theorem}
Hence, we obtain asymptotic valid tests $\varphi_{n,C} = I\{ S_{n,C} > \chi^2_{\text{rank}(\mbf{T}),1-\alpha}\}$ and $\varphi_{n,{B}} = \{ S_{n,{B}} > \chi^2_{\text{rank}(\mbf{T}),1-\alpha}\}$ for the null hypotheses \eqref{eqn:null} by comparing the respective WTS with the $(1-\alpha)$-quantile $\chi^2_{\text{rank}(\mbf{T}),1-\alpha}$ of a chi-square distribution with $\text{rank}(\mbf{T})$ degrees of freedom. 
As Theorem \ref{theo:conv_C+beta} is generally valid, we can deduce that $n^{-1}S_{n,C}(\mbf{T})$ converges always in probability to $\widetilde S_C=(\mbf{T}\mbf{C})\trans(\mbf{T}\mbf{\Sigma}_C\mbf{T}\trans)^+\mbf{T}\mbf{C}$. In the proofs, we verify that this limit $\widetilde S_C$ is positive whenever $\mbf{T}\mbf{C}\neq \mbf{0}$ holds. Consequently, the consistency of the tests follows:

\begin{theorem}\label{theo:WTS_alt}
	Suppose \assumption. Then $\varphi_{n,C}$ and $\varphi_{n,{B}}$ are consistent, i.e.
	\begin{enumerate}[(i)]
		\item\label{enu:theo:WTS_alt_C} $\E_{\mathcal H_{1,C}}(\varphi_{n,C})\to 1$ for $\mathcal H_{1,C}: \mbf{T}\mbf{C} \neq 0$.
		
		\item\label{enu:theo:WTS_alt_beta} $\E_{\mathcal H_{1,{B}}}(\varphi_{n,{B}})\to 1$ for $\mathcal H_{1,{B}}: \mbf{T}\mbf{{B}} \neq 0$.
	\end{enumerate}
\end{theorem}
It is well known that Wald-type statistics, as used here, converge rather slowly to their $\chi^2$-distributed limit. This explains the poor type-1 error control in our simulation study, see Section \ref{sec:simulation}, where diverse small sample size settings are considered. This problem can be tackled by the permutation method mentioned in the introduction and explained more detail in the following section.

\section{Permutation method}\label{sec:permutation} 
Resampling techniques and, in particular, permutation methods are useful and well-accepted tools to achieve better finite sample performance. The benefit of the permutation approach is its finite exactness under exchangeability, here under $\mathcal {\widetilde H}_0: \mbf{X}_{11} \overset{d}{=}\ldots \overset{ d}{=} \mbf{X}_{k1}$. At the same time, the tests' asymptotic properties, as being asymptotically exact under the null hypothesis and consistent under alternatives, can be transferred to the permutation counterpart when an appropriate studentization is used within the original test statistic. The WTS is a perfect example for such a studentized statistic.

Let us become more specific now. To generate a permutation sample $\mbf{X}^\pi=(\mbf{X}^\pi_{ij})_{i=1,\ldots,k;j=1,\ldots,n_i}$, we pool all the observations and forget the corresponding group memberships for a moment. For each group $j$, we draw then a new sample of its original size $n_j$ from the pooled data $\mbf{X}=(\mbf{X}_{ij})_{i=1,\ldots,k;j=1,\ldots,n_i}$, but, in contrast to Efron's bootstrap, we draw without replacement. In short, we randomly permute the group memberships. Replacing the original data by the permutation sample, we obtain the permutation counterparts $S_{n,C}^\pi(\mbf{T})$ and $S_{n,{B}}^\pi(\mbf{T})$ of the WTS. In same way, we add the superscript $^\pi$ to the variance estimators, the covariance estimators etc. when the corresponding permutation counterpart is meant. Since we pool the data for the permutation approach, we need to adjust the condition on the expectation vectors to prevent division by zero. To be concrete, we suppose that the expectation vector $\mbf{\bar\mu} = \sum_{i=1}^k\kappa_i\mbf{\mu}_i$ is not equal to zero. 
\begin{theorem}\label{theo:perm} Suppose \assumption{} and $\mbf{\bar\mu} \neq \mbf{0}$. Under $\mathcal H_{0,C}: \mbf{T}\mbf{C}=\mbf{0}$ as well as under $\mathcal H_{1,C}:\mbf{T}\mbf{C}\neq \mbf{0}$, $S_{n,C}^\pi(\mbf{T})$ always mimics the null distribution limit of $S_{n,C}(\mbf{T})$ asymptotically, i.e.
	\begin{align*}
	\sup_{x\in\R} \Big| \P\Big(S_{n,C}^\pi(\mbf{T}) \leq x \mid \mbf{X} \Big) - \chi^2_{\text{rank}(\mbf{T})}(x) \Big| \overset{p}{\rightarrow} 0,
	\end{align*}
	where $\chi^2_{\text{rank}(\mbf{T})}$ denotes the distribution function of a chi-square distribution with $\text{rank}(\mbf{T})$ degrees of freedom. The analog statement for $S_{n,{B}}^\pi(\mbf{T})$ is true.
\end{theorem}
By Theorem \ref{theo:perm}, the $(1-\alpha)$-quantile $q^\pi_{n\alpha,C}$ of the permutation distribution $t\mapsto \P(S_{n,C}^\pi(\mbf{T}) \leq t \mid \mbf{X} )$ can be used to approximate the chi-square quantile $\chi^2_{\text{rank}(\mbf{T}),1-\alpha}$, independently whether the null hypothesis or the alternative is true. The analog statement is true for testing in terms of the parameter $\mbf{{B}}$. Consequently, the asymptotic exactness and the consistency of the tests $\varphi_{n,C}$ and $\varphi_{n,{B}}$ can be transferred to their permutation counterparts $\varphi_{n,C}^\pi = \I\{ S_{n,C} > q^\pi_{n\alpha, C}\}$ and $\varphi_{n,{B}} = \I\{ S_{n,{B}} > q^\pi_{n\alpha,{B}}\}$ \citep[Lemma 1 and Theorem 7]{janssenPauls2003}.

As a small byproduct, Theorem \ref{theo:perm} shows that the permutation results of \cite{PaulySmaga2020} do not require the specific convergence rate assumption $n_i/n-\kappa_i = O(n^{-1/2})$ but hold in general under \eqref{eqn:ni/n}.

\section{Simulation study}
\label{sec:simulation}
In this section, we present the simulation study to investigate the type-1 error level and power of our new tests under small and moderate sample sizes for 
\begin{enumerate}[(i)]
	\item interfering the null hypothesis of equal $C$'s, and $B$'s, in multivariate one-way layouts for $d=5,10$.
	\item testing for the presence of main or interaction effects in univariate ($d=1$) two-way layouts.
\end{enumerate}
Therefore, we consider
\begin{enumerate}	
	\item different distributions: the normal ($N$), the power exponential ($\mathrm{PE}_2$, $\mathrm{PE}_{.5}$) and the Student ($t_5$) distributions with covariance matrix $\mbf{\Sigma}_i  = \mbf{I}_d$ and mean vector $\mbf{\mu}_{i1}=(1/C_i)\mbf{e}_1$ or $\mbf{\mu}_{i2}=(1/C_i)\mbf{1}_d/\sqrt{n}$, where $\mbf{e}_1=(1,\mbf{0}_{d-1}^{\top})^{\top}$. 
	
	\item balanced sample size settings, $n_i=n_0\in\{20,35,50\}$, as well as unbalanced scenarios, and
	
	\item different group sizes $k=2,4,8$.
\end{enumerate}
The univariate one-way layout was already explicitly studied by \cite{PaulySmaga2020} and we refer the reader to their paper for comprehensive simulations within this context. As competitors in the $k$-sample settings, we choose the different tests suggested by \cite{AertsHaesbroeck2017}, i.e., the asymptotic $AH_{C}$, $AH_{R}$, $AH_{S}$ and $AH_{SP}$ tests based on the classical, one-step reweighted minimum covariance determinant (RMCD), S estimators and semi-parametric approach, respectively. For all these tests there is version based on $C$ and $B$, respectively. To indicate the specific version, we add the index $C$ and $B$, respectively, e.g. $AH_{C,C}$ and $AH_{C,B}$, when this is necessary. The robust estimators (RMCD, S) were performed with a default breakdown point (BDP) of $25\%$. In contrast to our approach, the tests of \cite{AertsHaesbroeck2017} are  (semi-)parametric and information about the underlying distribution is needed. Although this knowledge is typically not given in practice, the asymptotic variances of and consistency factors for the estimators (classical, S, RMCD) were chosen under the assumed underlying distributions.

The significance level was set to $\alpha=5\%$. Empirical sizes and powers of the tests were computed as the proportion of rejections of the null hypothesis based on $1000$ simulation replications. The $p$-values of the permutation tests were estimated by using $1000$ random permutations. The simulation experiments were performed using the R program \citep{Rcore}. The code for the tests proposed in \cite{AertsHaesbroeck2017} was taken from the 
ResearchGate profile of Doctor Stephanie Aerts. A part of calculations was made at the Pozna\'n Supercomputing and Networking Center.

\subsection{Multivariate one-way layouts}
For the multivariate one-way layout, we divide our simulations into two parts. First, we consider balanced settings for two-group comparisons following the simulations settings of \cite{AertsHaesbroeck2017}, which include the choice $\mbf{\mu}_i=\mbf{\mu}_{i1}=(1/C_i)\mbf{e}_i$ for the mean vectors and $d=5$ for the number of dimensions. Therein, we investigate the type-1 error for different scenarios $C_1=C_2\in\{0.1,0.5,1,2\}$  as well as the power values under various settings, $C_1=1$ and $C_2\in\{0.5,1.5\}$, of our new tests and of the ones suggested by \cite{AertsHaesbroeck2017}. We want to point out that in one-way layouts both null hypotheses $\mathcal H_{0,C}$ and $\mathcal H_{0,B}$ are equivalent. That is why both versions of each test are include in all the corresponding simulations. The results are displayed in Tables \ref{table2} and \ref{table3}. Second, we run extra simulations for an in-depth analysis of our tests' finite sample performance. The observed values are presented in Tables \ref{table4} and \ref{table5}. They cover two-group but also four-group comparisons, another choice for the mean vector $\mbf{\mu}_i=\mbf{\mu}_{i1}=(1/C_i)\mbf{e}_i$, an unbalanced sample size setting $(n_1=35,n_2=45,n_3=40,n_4=50)$ and the dimensions $d=5,10$. Alternatives are chosen such that only the MCV from group $1$ differs from the remaining ones, i.e. $A: C_1 \neq C_2 = \ldots =C_k$.	

\begin{sidewaystable}
	\centering
	\caption{Empirical sizes (in $\%$) of the new asymptotic (Asy) and permutation (Per) tests as well as the tests by \cite{AertsHaesbroeck2017} ($AH_{C}$, $AH_{R}$, $AH_{S}$, $AH_{SP}$) under various balanced two-group settings}\label{table2}\scriptsize
	\scriptsize
	\begin{tabular}{cc|ccc|ccc|ccc|ccc|ccc|ccc|}
		\hline
		&$C_0$&&\multicolumn{4}{c}{0.1}& && \multicolumn{4}{c}{0.5} && &\multicolumn{4}{c}{2}&\\ 
		&& &$C$&& &$B$& &&$C$&& &$B$&&&$C$&&&$B$&\\
		&$n_0$& 20&35&50&20&35&50 & 20&35&50&20&35&50 & 20&35&50&20&35&50\\
		\hline
		$\mathrm{PE}_2$&$\text{Asy}$&{\bf 12.5}&{\bf 9.0}&{\bf 6.5}&{\bf 12.4}&{\bf 8.5}&6.0 & {\bf 10.3}&6.3&5.1&{\bf 11.1}&{\bf 7.4}&5.2 & {\bf 0.2}&{\bf 0.1}&{\bf 0.2}&{\bf 7.0}&4.3&4.3\\
		&$\text{Per}$&5.2&5.7&5.0&5.3&5.1&5.0 & 5.3&5.1&4.5&5.6&5.0&4.3 & 5.3&4.2&5.0&5.0&4.6&4.6\\
		&$AH_{C}$& {\bf 9.5}&{\bf 7.7}&5.7&{\bf 9.5}&{\bf 7.7}&5.7 & {\bf 6.8}&6.1&{\bf 7.5}&{\bf 8.4}&{\bf 7.3}&{\bf 8.5} & {\bf 0.1}&{\bf 0.1}&{\bf 0.0}&{\bf 7.3}&6.1&5.1\\
		&$AH_{R}$& {\bf 27.7}&{\bf 25.4}&{\bf 17.7}&{\bf 27.9}&{\bf 25.4}&{\bf 17.7} & {\bf 27.3}&{\bf 24.5}&{\bf 18.7}&{\bf 30.6}&{\bf 26.1}&{\bf 20.1} & 4.8&4.8&{\bf 3.1}&{\bf 24.4}&{\bf 22.8}&{\bf 18.0}\\
		&$AH_{S}$& {\bf 8.7}&{\bf 7.3}&6.0&{\bf 8.8}&{\bf 7.4}&6.0 & 6.4&5.2&{\bf 7.2}&{\bf 8.8}&6.2&{\bf 8.3} & {\bf 0.1}&{\bf 0.1}&{\bf 0.1}&{\bf 7.3}&5.9&5.2\\
		&$AH_{SP}$& {\bf 10.1}&{\bf 7.8}&6.0&{\bf 10.4}&{\bf 7.7}&6.0 & {\bf 7.2}&6.1&{\bf 7.5}&{\bf 8.6}&{\bf 7.3}&{\bf 8.6} & {\bf 0.1}&{\bf 0.1}&{\bf 0.0}&{\bf 7.3}&6.2&5.1\\
		\hline
		$N$&$\text{Asy}$&{\bf 12.4}&{\bf 7.8}&{\bf 7.3}&{\bf 11.6}&{\bf 7.9}&{\bf 6.9} & {\bf 9.7}&{\bf 6.5}&{\bf 6.6}&{\bf 11.9}&{\bf 7.4}&{\bf 7.1} & {\bf 0.2}&{\bf 0.4}&{\bf 0.2}&{\bf 7.8}&4.8&5.6\\
		&$\text{Per}$&5.6&4.0&4.0&5.7&4.1&4.0 & 4.5&4.6&4.2&4.5&4.9&4.4 & 6.2&4.1&5.7&4.4&4.2&5.7\\
		&$AH_{C}$& {\bf 8.6}&4.8&5.8&{\bf 8.6}&4.9&5.9 & {\bf 7.0}&5.5&4.8&{\bf 9.1}&{\bf 6.8}&6.0 & {\bf 0.0}&{\bf 0.1}&{\bf 0.0}&{\bf 6.7}&4.8&4.9\\
		&$AH_{R}$& {\bf 21.7}&{\bf 15.3}&{\bf 10.6}&{\bf 21.8}&{\bf 15.5}&{\bf 10.9} & {\bf 23.3}&{\bf 13.6}&{\bf 9.0}&{\bf 26.3}&{\bf 15.4}&{\bf 10.1} & 5.0&{\bf 1.8}&{\bf 0.6}&{\bf 21.0}&{\bf 13.6}&{\bf 9.4}\\
		&$AH_{S}$& {\bf 8.4}&5.1&5.9&{\bf 8.5}&5.1&5.9 & {\bf 6.8}&5.6&4.9&{\bf 9.4}&{\bf 6.5}&6.0 & {\bf 0.0}&{\bf 0.1}&{\bf 0.0}&{\bf 6.5}&4.7&5.2\\
		&$AH_{SP}$& {\bf 11.0}&{\bf 6.9}&{\bf 6.7}&{\bf 11.5}&{\bf 7.3}&{\bf 6.8} & {\bf 8.4}&6.3&5.6&{\bf 11.0}&{\bf 9.0}&6.2 & {\bf 0.1}&{\bf 0.1}&{\bf 0.0}&{\bf 7.1}&5.1&5.3\\
		\hline
		$\mathrm{PE}_{.5}$&$\text{Asy}$&{\bf 15.2}&{\bf 10.8}&{\bf 9.5}&{\bf 16.0}&{\bf 11.5}&{\bf 10.4} & {\bf 11.7}&{\bf 10.2}&{\bf 6.7}&{\bf 15.3}&{\bf 12.5}&{\bf 8.8} & {\bf 1.0}&{\bf 0.2}&{\bf 0.2}&{\bf 8.4}&{\bf 6.6}&6.3\\
		&$\text{Per}$&5.4&5.1&4.9&5.1&4.6&4.9 & 4.7&6.0&4.3&5.6&5.5&4.4 & 4.7&5.0&5.5&4.4&5.1&5.4\\
		&$AH_{C}$& {\bf 7.2}&5.2&4.3&{\bf 7.2}&5.2&4.4 & 4.3&4.2&5.0&5.8&5.2&5.7 & {\bf 0.0}&{\bf 0.0}&{\bf 0.0}&4.8&4.6&3.9\\
		&$AH_{R}$& {\bf 16.3}&{\bf 8.5}&6.0&{\bf 16.3}&{\bf 8.4}&5.9 & {\bf 12.7}&{\bf 8.1}&{\bf 7.8}&{\bf 15.1}&{\bf 9.2}&{\bf 8.5} & {\bf 1.4}&{\bf 0.3}&{\bf 0.1}&{\bf 12.3}&{\bf 6.6}&4.8\\
		&$AH_{S}$& {\bf 8.6}&{\bf 6.6}&5.1&{\bf 8.6}&{\bf 6.6}&5.1 & 5.3&5.1&5.2&{\bf 7.0}&{\bf 6.5}&6.3 & {\bf 0.1}&{\bf 0.0}&{\bf 0.0}&5.2&4.2&4.1\\
		&$AH_{SP}$& {\bf 14.2}&{\bf 8.3}&{\bf 6.6}&{\bf 14.1}&{\bf 8.4}&{\bf 7.0} & {\bf 9.3}&{\bf 6.6}&{\bf 6.7}&{\bf 11.1}&{\bf 9.2}&{\bf 7.7} & {\bf 0.0}&{\bf 0.0}&{\bf 0.0}&{\bf 7.8}&5.2&4.2\\
		\hline
		$t_5$&$\text{Asy}$&{\bf 17.7}&{\bf 13.5}&{\bf 11.8}&{\bf 19.8}&{\bf 14.6}&{\bf 13.0} & {\bf 14.5}&{\bf 11.2}&{\bf 9.6}&{\bf 17.9}&{\bf 15.1}&{\bf 11.2} & {\bf 1.3}&{\bf 0.3}&{\bf 0.1}&{\bf 7.7}&{\bf 9.2}&{\bf 6.9}\\
		&$\text{Per}$&5.3&5.6&5.1&5.4&4.9&5.2 & 5.0&5.2&4.6&4.9&5.7&4.8 & 5.2&5.1&5.7&4.2&5.8&5.6\\
		&$AH_{C}$& {\bf 0.1}&{\bf 0.3}&{\bf 0.7}&{\bf 0.1}&{\bf 0.3}&{\bf 0.7} & {\bf 0.1}&{\bf 0.2}&{\bf 1.0}&{\bf 0.3}&{\bf 0.2}&{\bf 1.3} & {\bf 0.0}&{\bf 0.0}&{\bf 0.0}&{\bf 0.8}&{\bf 1.1}&{\bf 0.9}\\
		&$AH_{R}$& {\bf 9.8}&5.8&5.9&{\bf 9.9}&5.8&6.0 & {\bf 7.2}&5.2&4.2&{\bf 8.5}&6.2&4.8 & {\bf 0.9}&{\bf 0.0}&{\bf 0.0}&{\bf 7.9}&4.9&3.9\\
		&$AH_{S}$& {\bf 7.4}&6.1&5.7&{\bf 7.4}&6.1&5.7 & 5.1&5.3&4.7&{\bf 7.4}&6.3&5.5 & {\bf 0.2}&{\bf 0.0}&{\bf 0.1}&6.1&4.7&{\bf 3.2}\\
		&$AH_{SP}$& {\bf 14.0}&{\bf 10.4}&{\bf 9.2}&{\bf 15.2}&{\bf 11.3}&{\bf 9.8} & {\bf 9.0}&{\bf 8.3}&6.4&{\bf 11.9}&{\bf 10.2}&{\bf 8.1} & {\bf 0.1}&{\bf 0.0}&{\bf 0.0}&{\bf 7.2}&5.1&4.0\\
		\hline
	\end{tabular}

	Here, $\mathcal H_{0,C}:C_1=C_2=C_0$, $n_1=n_2=n_0$, $C$ and $B$ denote that the tests for MCVs and for inverses of MCVs are used respectively. The empirical sizes are displayed in bold, when they are outside the $95\%$ significance limits, i.e., $[3.6\%,6.4\%]$.
\end{sidewaystable}

\begin{table}
	\centering
	\centering
	\caption{Empirical sizes ($C_2=1$) and powers ($C_2=0.5,1.5$) (in $\%$) of the new asymptotic (Asy) and permutation (Per) tests as well as the tests by \cite{AertsHaesbroeck2017} ($AH_{C}$, $AH_{R}$, $AH_{S}$, $AH_{SP}$) under balanced two-group settings}\label{table3}\scriptsize
	\scriptsize
	\begin{tabular}{cc|cc|cc|cc|}
		\hline
		&$C_2$& \multicolumn{2}{c}{0.5} & \multicolumn{2}{c}{1} & \multicolumn{2}{c}{1.5}\\ 
		Distr&Test& $C$&$B$ & $C$&$B$ & $C$&$B$\\
		\hline
		$\mathrm{PE}_2$&$\text{Asy}$&90.6&{\bf 93.5}&4.4&{\bf 7.2}&13.0&{\bf 29.7}\\
		&$\text{Per}$&90.1&91.7&6.1&5.6&26.0&27.6\\
		&$AH_{C}$& 93.4& 94.5& {\bf 3.1}& 6.4& 10.5& 29.2\\
		&$AH_{R}$& {\bf 74.8}& {\bf 78.0}& {\bf 14.4}& {\bf 20.4}& {\bf 17.5}& {\bf 34.3}\\
		&$AH_{S}$& 91.0& 93.3& {\bf 3.0}& 6.2& 7.5& 27.4\\
		&$AH_{SP}$& 93.4& {\bf 94.6}& {\bf 3.2}& {\bf 6.7}& 10.5& {\bf 29.1}\\
		\hline
		$N$&$\text{Asy}$&89.4&92.4&{\bf 3.1}&6.1&12.6&30.4\\
		&$\text{Per}$&88.4&89.0&4.4&4.7&25.2&26.1\\
		&$AH_{C}$& 88.5& {\bf 92.0}& 3.6& {\bf 7.7}& 13.5& {\bf 29.1}\\
		&$AH_{R}$& {\bf 79.5}& {\bf 83.5}& {\bf 7.3}& {\bf 12.3}& {\bf 13.9}& {\bf 28.5}\\
		&$AH_{S}$& 87.0& {\bf 90.1}& {\bf 3.5}& {\bf 7.2}& 12.3& {\bf 28.6}\\
		&$AH_{SP}$& 89.1& {\bf 92.0}& 4.1& {\bf 7.9}& 14.1& {\bf 30.0}\\
		\hline
		$\mathrm{PE}_{.5}$&$\text{Asy}$&87.6&{\bf 89.5}&4.5&{\bf 8.3}&12.7&{\bf 29.1}\\
		&$\text{Per}$&85.8&84.5&5.5&6.0&22.7&23.5\\
		&$AH_{C}$& 83.8& 86.6& {\bf 2.0}& 5.0& 6.8& 22.9\\
		&$AH_{R}$& 79.1& 83.5& {\bf 3.4}& 6.3& 9.7& 24.6\\
		&$AH_{S}$& 86.7& 89.1& {\bf 2.9}& 5.5& 10.5& 24.6\\
		&$AH_{SP}$& 86.4& 88.8& {\bf 2.5}& 6.3& 9.4& 25.8\\
		\hline
		$t_5$&$\text{Asy}$&82.6&{\bf 84.6}&5.9&{\bf 10.6}&18.7&{\bf 33.3}\\
		&$\text{Per}$&78.7&76.1&5.9&5.3&25.2&24.9\\
		&$AH_{C}$& 42.9& 52.5& {\bf 0.1}& {\bf 1.4}& 0.8& 12.8\\
		&$AH_{R}$& 77.5& 81.5& {\bf 2.5}& 4.8& 12.0& 25.4\\
		&$AH_{S}$& 81.8& 84.4& {\bf 3.0}& 5.8& 15.4& 30.3\\
		&$AH_{SP}$& 78.6& {\bf 81.9}& 4.2& {\bf 8.4}& 12.3& {\bf 28.8}\\
		\hline
	\end{tabular}

	Here, $\mathcal H_{0,C}:C_1=C_2$, $C_1=1$, $n_1=n_2=50$, $C$ and $B$ denote that the tests for MCVs and for inverses of MCVs are used respectively. The empirical sizes are displayed in bold, when they are outside the $95\%$ significance limits, i.e., $[3.6\%,6.4\%]$. The empirical powers are in bold for too liberal tests, i.e., when the corresponding empirical sizes are greater then $6.4\%$.
\end{table}

\begin{table}
	\centering
	\caption{Empirical sizes (in $\%$) of the new asymptotic ($\text{Asy}$) and permutation ($\text{Per}$) tests under the unbalanced setting $(n_1=35,n_2=45,n_3=40,n_4=50)$ for the expectation vectors $\mbf{\mu}_i=\mbf{\mu}_{i2}$}\label{table4}\scriptsize
	\scriptsize
	\begin{tabular}{cccc|cc|cc|cc|cc|cc|}
		\hline
		&&&$C_0$&\multicolumn{2}{c}{0.1}&\multicolumn{2}{c}{0.5}&\multicolumn{2}{c}{1}&\multicolumn{2}{c}{1.5}&\multicolumn{2}{c}{2}\\ 
		Distr&$k$&$d$&Test& $C$&$B$&$C$&$B$&$C$&$B$&$C$&$B$&$C$&$B$\\
		\hline
		$\mathrm{PE}_2$&2&5&$\text{Asy}$& {\bf 7.4}&{\bf 7.0}&{\bf 6.7}&{\bf 6.8}&{\bf 3.4}&5.7&{\bf 1.1}&4.8&{\bf 0.4}&4.4\\
		&&&$\text{Per}$& 5.1&4.9&5.3&5.1&4.3&4.7&4.2&4.3&4.3&4.4\\
		&&10&$\text{Asy}$& {\bf 12.9}&{\bf 12.1}&{\bf 9.6}&{\bf 10.4}&{\bf 7.6}&{\bf 9.8}&3.9&{\bf 8.0}&{\bf 1.7}&{\bf 8.0}\\
		&&&$\text{Per}$& 6.0&5.7&4.6&4.9&5.3&5.2&4.7&4.3&5.0&5.1\\
		&4&5&$\text{Asy}$& {\bf 12.1}&{\bf 10.5}&{\bf 7.6}&{\bf 7.4}&4.3&{\bf 7.5}&{\bf 2.3}&{\bf 6.7}&{\bf 0.5}&5.6\\
		&&&$\text{Per}$& 5.8&5.8&4.0&4.9&4.0&5.1&4.1&5.1&5.5&4.9\\
		&&10&$\text{Asy}$& {\bf 16.7}&{\bf 14.9}&{\bf 14.4}&{\bf 14.7}&{\bf 11.0}&{\bf 14.8}&4.1&{\bf 10.1}&5.9&{\bf 12.2}\\
		&&&$\text{Per}$& 5.1&5.0&4.5&4.4&5.9&4.9&5.1&5.5&4.7&5.3\\
		\hline
		$N$&2&5&$\text{Asy}$& {\bf 9.5}&{\bf 8.9}&{\bf 6.5}&{\bf 7.1}&4.2&{\bf 6.9}&{\bf 1.2}&{\bf 6.5}&{\bf 0.3}&5.4\\
		&&&$\text{Per}$& 5.2&5.0&4.0&4.0&5.0&4.7&5.2&5.0&4.6&4.9\\
		&&10&$\text{Asy}$& {\bf 13.6}&{\bf 12.8}&{\bf 10.5}&{\bf 11.1}&{\bf 8.4}&{\bf 11.5}&{\bf 3.1}&{\bf 7.4}&{\bf 1.8}&{\bf 9.3}\\
		&&&$\text{Per}$& 5.8&6.2&4.8&4.4&6.0&6.1&4.0&4.2&5.2&5.5\\
		&4&5&$\text{Asy}$& {\bf 14.2}&{\bf 13.2}&{\bf 12.1}&{\bf 11.8}&6.4&{\bf 9.8}&{\bf 3.0}&{\bf 7.4}&{\bf 0.7}&5.8\\
		&&&$\text{Per}$& 5.3&5.4&5.9&5.4&5.3&4.8&5.0&5.0&4.5&4.4\\
		&&10&$\text{Asy}$& {\bf 16.7}&{\bf 15.4}&{\bf 19.0}&{\bf 20.0}&{\bf 9.7}&{\bf 13.3}&5.5&{\bf 11.6}&4.8&{\bf 10.3}\\
		&&&$\text{Per}$& 5.2&4.4&5.5&4.4&4.8&5.3&4.6&4.5&5.6&5.7\\
		\hline
		$\mathrm{PE}_{.5}$&2&5&$\text{Asy}$& {\bf 12.7}&{\bf 13.7}&{\bf 9.1}&{\bf 11.5}&4.2&{\bf 9.5}&{\bf 1.3}&{\bf 8.9}&{\bf 0.6}&{\bf 8.5}\\
		&&&$\text{Per}$& 5.9&6.0&5.0&4.6&4.5&4.5&5.6&5.3&6.3&5.8\\
		&&10&$\text{Asy}$& {\bf 14.4}&{\bf 13.5}&{\bf 12.4}&{\bf 12.6}&{\bf 8.1}&{\bf 11.8}&4.3&{\bf 10.6}&{\bf 3.0}&{\bf 10.2}\\
		&&&$\text{Per}$& 4.8&4.6&4.7&4.8&4.7&5.4&4.6&5.3&5.8&5.1\\
		&4&5&$\text{Asy}$& {\bf 17.7}&{\bf 18.7}&{\bf 13.7}&{\bf 16.0}&{\bf 7.6}&{\bf 12.4}&3.8&{\bf 10.0}&{\bf 1.3}&{\bf 8.7}\\
		&&&$\text{Per}$& 4.4&4.7&4.5&5.1&5.2&5.6&5.3&5.2&5.0&5.7\\
		&&10&$\text{Asy}$& {\bf 24.2}&{\bf 22.2}&{\bf 21.0}&{\bf 21.5}&{\bf 14.5}&{\bf 19.0}&5.6&{\bf 11.4}&{\bf 8.4}&{\bf 12.8}\\
		&&&$\text{Per}$& 5.0&4.8&4.9&4.1&6.3&5.9&5.5&5.5&5.0&4.4\\
		\hline
		$t_5$&2&5&$\text{Asy}$& {\bf 13.1}&{\bf 14.1}&{\bf 13.1}&{\bf 14.1}&{\bf 7.0}&{\bf 11.4}&{\bf 2.8}&{\bf 9.3}&{\bf 0.7}&{\bf 7.9}\\
		&&&$\text{Per}$& 4.3&4.8&4.3&4.8&5.0&5.0&5.4&5.2&5.8&4.7\\
		&&10&$\text{Asy}$& {\bf 20.7}&{\bf 19.9}&{\bf 17.4}&{\bf 18.2}&{\bf 11.1}&{\bf 14.0}&6.1&{\bf 11.9}&4.1&{\bf 10.7}\\
		&&&$\text{Per}$& 4.8&4.2&5.2&5.6&5.1&5.2&4.5&4.8&5.7&5.4\\
		&4&5&$\text{Asy}$& {\bf 22.7}&{\bf 24.6}&{\bf 17.3}&{\bf 20.2}&{\bf 10.7}&{\bf 16.1}&3.6&{\bf 10.9}&{\bf 1.6}&{\bf 9.0}\\
		&&&$\text{Per}$& 5.0&5.6&5.0&5.6&5.9&5.3&4.1&4.3&4.4&5.2\\
		&&10&$\text{Asy}$& {\bf 31.1}&{\bf 32.0}&{\bf 24.7}&{\bf 26.5}&{\bf 16.8}&{\bf 22.6}&6.1&{\bf 13.9}&{\bf 8.5}&{\bf 15.6}\\
		&&&$\text{Per}$& 5.0&4.4&4.7&5.3&4.1&4.9&5.5&6.1&4.2&4.5\\
		\hline
	\end{tabular}

	Here, $\mathcal H_{0,C}:C_1=\cdots=C_k=C_0$ and $C$ and $B$ denote that the tests for MCVs and for inverses of MCVs are used respectively. The empirical sizes are displayed in bold, when they are outside the $95\%$ significance limits, i.e., $[3.6\%,6.4\%]$.
\end{table}

\begin{table}
	\centering
	\caption{Empirical powers (in $\%$) of the new asymptotic ($\text{Asy}$) and permutation ($\text{Per}$) tests under alternatives 
		$A: C_1 \neq C_2 = \ldots =C_k$ for the unbalanced setting $(n_1=35,n_2=45,n_3=40,n_4=50)$ and the expectation vectors $\mbf{\mu}_i=\mbf{\mu}_{i2}$}\label{table5}\scriptsize
	\scriptsize
	\begin{tabular}{cccc|cc|cc|cc|cc|}
		\hline
		&&&$(C_1,C_2)$&\multicolumn{2}{c}{$(0.07, 0.1)$}& \multicolumn{2}{c}{$(0.13,0.1)$}&\multicolumn{2}{c}{$(0.5,1)$}&\multicolumn{2}{c}{$(1.5,1)$}\\ 
		Distr&$k$&$d$&Test& $C$&$B$&$C$&$B$&$C$&$B$&$C$&$B$\\
		\hline
		$\mathrm{PE}_2$&2&5&$\text{Asy}$& {\bf 72.3}&{\bf 70.3}&{\bf 37.8}&{\bf 40.4}&{\bf 87.0}&{\bf 89.9}&2.3&20.5\\
		&&&$\text{Per}$& 70.9&71.1&30.1&35.0&85.4&85.2&8.1&17.0\\
		&&10&$\text{Asy}$& {\bf 71.9}&{\bf 70.7}&{\bf 29.4}&{\bf 30.7}&{\bf 87.8}&{\bf 88.7}&{\bf 4.8}&{\bf 15.9}\\
		&&&$\text{Per}$& 62.1&63.9&17.5&20.0&79.6&79.9&4.3&9.2\\
		&4&5&$\text{Asy}$& {\bf 79.7}&{\bf 68.4}&{\bf 38.3}&{\bf 44.0}&{\bf 94.7}&{\bf 89.3}&5.9&{\bf 21.6}\\
		&&&$\text{Per}$& 61.7&59.0&16.9&29.0&92.7&82.8&6.2&15.1\\
		&&10&$\text{Asy}$& {\bf 81.6}&{\bf 73.1}&{\bf 33.3}&{\bf 37.8}&{\bf 93.8}&{\bf 90.5}&{\bf 8.2}&{\bf 18.8}\\
		&&&$\text{Per}$& 56.2&51.4&8.8&17.1&86.4&72.2&4.3&7.2\\
		\hline
		$N$&2&5&$\text{Asy}$& {\bf 66.7}&{\bf 65.5}&{\bf 33.5}&{\bf 35.7}&{\bf 83.2}&{\bf 85.8}&4.8&{\bf 20.6}\\
		&&&$\text{Per}$& 63.6&64.6&24.8&29.3&81.4&80.7&10.3&17.0\\
		&&10&$\text{Asy}$& {\bf 70.2}&{\bf 68.9}&{\bf 31.0}&{\bf 32.6}&{\bf 86.6}&{\bf 88.6}&{\bf 6.6}&{\bf 17.4}\\
		&&&$\text{Per}$& 60.3&62.2&16.5&20.6&78.7&75.7&4.6&9.2\\
		&4&5&$\text{Asy}$& {\bf 71.5}&{\bf 60.3}&{\bf 35.5}&{\bf 41.2}&{\bf 92.0}&{\bf 87.2}&7.0&{\bf 26.1}\\
		&&&$\text{Per}$& 54.1&52.8&13.5&25.7&89.1&79.0&6.5&17.8\\
		&&10&$\text{Asy}$& {\bf 78.7}&{\bf 69.6}&{\bf 33.1}&{\bf 37.5}&{\bf 93.8}&{\bf 88.7}&{\bf 9.7}&{\bf 22.5}\\
		&&&$\text{Per}$& 52.6&50.2&9.7&17.1&83.0&72.5&4.9&9.1\\
		\hline
		$\mathrm{PE}_{.5}$&2&5&$\text{Asy}$& {\bf 59.8}&{\bf 57.2}&{\bf 31.4}&{\bf 34.8}&{\bf 80.9}&{\bf 83.0}&3.8&{\bf 20.0}\\
		&&&$\text{Per}$& 55.4&54.7&21.5&26.0&77.0&73.5&7.3&13.5\\
		&&10&$\text{Asy}$& {\bf 66.0}&{\bf 65.0}&{\bf 30.0}&{\bf 32.9}&{\bf 85.2}&{\bf 85.8}&{\bf 7.7}&{\bf 19.8}\\
		&&&$\text{Per}$& 55.4&57.7&17.7&20.3&75.9&74.4&5.4&9.5\\
		&4&5&$\text{Asy}$& {\bf 66.6}&{\bf 57.3}&{\bf 35.1}&{\bf 39.4}&{\bf 86.0}&{\bf 80.9}&{\bf 9.8}&{\bf 27.2}\\
		&&&$\text{Per}$& 51.3&51.0&13.5&24.2&80.0&66.5&7.5&14.3\\
		&&10&$\text{Asy}$& {\bf 75.3}&{\bf 68.9}&{\bf 36.6}&{\bf 41.4}&{\bf 92.0}&{\bf 88.0}&{\bf 10.7}&{\bf 22.5}\\
		&&&$\text{Per}$& 51.2&48.7&9.6&18.3&82.1&70.0&4.9&7.6\\
		\hline
		$t_5$&2&5&$\text{Asy}$& {\bf 55.1}&{\bf 54.3}&{\bf 31.2}&{\bf 33.4}&{\bf 82.9}&{\bf 84.0}&{\bf 7.1}&{\bf 24.8}\\
		&&&$\text{Per}$& 52.5&53.5&22.2&26.1&78.1&73.8&8.7&15.4\\
		&&10&$\text{Asy}$& {\bf 65.5}&{\bf 64.1}&{\bf 32.2}&{\bf 33.7}&{\bf 86.0}&{\bf 86.5}&{\bf 11.3}&{\bf 23.5}\\
		&&&$\text{Per}$& 56.2&57.7&17.8&22.6&74.3&74.8&6.4&11.8\\
		&4&5&$\text{Asy}$& {\bf 64.7}&{\bf 61.4}&{\bf 39.9}&{\bf 45.8}&{\bf 84.6}&{\bf 81.4}&{\bf 10.6}&{\bf 32.1}\\
		&&&$\text{Per}$& 51.9&55.2&17.8&28.7&78.0&63.6&7.2&15.9\\
		&&10&$\text{Asy}$& {\bf 76.9}&{\bf 73.7}&{\bf 42.8}&{\bf 47.8}&{\bf 90.4}&{\bf 88.7}&{\bf 18.3}&{\bf 32.9}\\
		&&&$\text{Per}$& 55.2&58.3&13.1&22.8&80.4&67.7&6.9&11.5\\
		\hline
	\end{tabular}

	Here, $C$ and $B$ denote that the tests for MCVs and for inverses of MCVs are used respectively. The empirical powers are in bold for too liberal tests, i.e., when the corresponding empirical sizes are greater then $6.4\%$.
\end{table}

\textit{Type-1 error:} As well-known in the literature, Wald-type statistics converge rather slowly to their chi-square limit distribution leading to a poor type-1 error control of the corresponding tests. This unsatisfactory behavior can also be observed in the present situation, see Tables \ref{table2}--\ref{table5}. Both asymptotic tests $\varphi_{n,C}$ and $\varphi_{n,B}$ fail to maintain the type-1 error level in most scenarios. The $95\%$ binomial confidence interval $[3.6\%,6.4\%]$ for the true type-1-error probability is used as criteria \citep{DuchesneFrancq2015} for liberality and conservativeness. The test $\varphi_{n,B}$ leads to rather liberal decisions with values even up to $32\%$ (Table \ref{table4}, $d=5$, $k=4$, $t_5$-distr.). In contrast to that, $\varphi_{n,C}$ exhibit extreme liberal but also extreme conservative decisions with observed type-1 values ranging from $0.1\%$ (Table \ref{table2}, several times) to $31.1\%$ (Table \ref{table4}, $d=5$, $k=4$, $t_5$-distr.). In principle, it can be observed that too liberal decision of $\varphi_{n,C}$ can be observed for small and moderate MCVs (i.e., $C_i=0.1,0.5$) and too conservative type-1 error values appear in case of large MCVs $(C_i=1.5,2)$. Table \ref{table2} shows that the empirical sizes come closer to the $5\%$-benchmark for increasing sample size, except $\varphi_{n,C}$ for $C=2$, which remains extremely conservative. Moreover, according to Table \ref{table4} the decisions of both tests become more liberal when the number $k$ of groups or dimension $d$ grows. \\
The competing procedures of \cite{AertsHaesbroeck2017} lead, in principle, to less liberal decisions for small and moderate $C_i=0.1,0.5$ than our asymptotic tests, but they are still too liberal, in particular for small sample sizes. For larger MCV ($C_i=2$), the type-1 error rates of the $AH$ tests and our asymptotic tests are comparable, i.e. the tests based on $C$ are extremely conservative and the tests based on $B$ are rather liberal. The $AH_R$ test makes an exception from all these observations under the normal distribution and the power exponential distribution $PE_2$: its liberality is even more pronounced for small and moderate $C_i=0.1,0.5$ while in case of $C_i=2$ the empirical sizes of $AH_{R,C}$ are less conservative than the ones of the other tests based on $C$. The overall impression is that $AH$ procedure exhibit a better type-1 error control than our asymptotic tests. This may be explained by the fact that these tests use more information, i.e. they require the knowledge of the underlying distribution, than our nonparametric approach. \\
As described before, the type-1 error control of $AH$ and our asymptotic tests is unsatisfactory, including too liberal but also too conservative decisions. In contrast, the permutation counterparts $\varphi_{n,C}^\pi$ and $\varphi_{n,B}^\pi$ of our asymptotic tests control the type-1 error rate very accurately over all settings. The corresponding empirical sizes lie always in the $95\%$ binomial confidence interval.

\textit{Power:} Due to the partially quite liberal behavior of the $AH$ and our asymptotic tests, a comparison of them and the permutation tests is not fair. Nevertheless, the empirical power values of all tests are presented in Tables \ref{table3} and \ref{table5} for completeness reasons. But the values corresponding to liberal tests (i.e. the empirical sizes is greater than the upper bound $6.4\%$ of the confidence interval) are displayed in bold to minimize the risk of misinterpretation. In the balanced settings (see Table \ref{table3}) the power values of the permutation test are comparable or slightly smaller than the other tests for small and moderate $C_i=0.1,0.5$, while the permutation test $\varphi_{n,C}^\pi$ clearly outperforms the others tests based on $C$ in case of $C_i=2$. The latter can be explained by the observed extreme conservativeness of the $AH$ and our asymptotic tests.  The $AH_R$ tests make again an exception. Here, we can observe that despite its extreme liberality under PE$_2$- and normal distributions in case of $C_i=0.5$, the corresponding power is significantly smaller than the one of the other tests. This may suggest some instability using the RMCD estimator with the asymptotic distribution. In the unbalanced settings (Table \ref{table5}) the empirical power values of the asymptotic and the permutation tests are close together for $k=2$ and $d=5$, where there is a slight advantage of the asymptotic tests. When switching to $k=4$ and/or $d=10$, the gap between the permutation and asymptotic tests in terms of power becomes larger and larger. While the power remains stable or decreases slightly - the natural power behavior for increasing dimensions - the power of the asymptotic tests even increases leading to the aforementioned growing gap. Our findings regarding the type-1 error control, i.e. increasing empirical size for growing dimension, explain this rather unusual power behavior of the asymptotic tests.

\textbf{Recommendation:} Summarizing the findings, we recommend the use of the permutation methods. They exhibit an accurate type-1 error control while the $AH$ and our asymptotic tests are rather unstable, including extreme liberal as well as conservative decisions. Moreover, the power performance is comparable or even significantly better in the situations, in which a fair comparison can be made.

\subsection{Two-way layouts}\label{sec:simu_fac+design}
In this section, we consider univariate ($d=1$) two way-layouts with factor $A$ possessing $a$ and factor $B$ having $b$ levels. Before we explain the specific simulation settings let us explain how to formulate the null hypotheses to check for main or interaction effects. For ease of presentation, we just discuss the null hypotheses and tests based on $C$ but the same can be done analogously for standardized means $B$.

First, recall the definitions $\mbf{J}_r = \mbf{1}_r\mbf{1}_r\trans$ and $\mbf{P}_r = \mbf{I}_r - \mbf{J}_r/r$ for $r\in \N$. Then the null hypotheses of interest can be described by the following contrast matrices $\mathcal H$:
\begin{itemize}
	\item  \textit{No main effect A}:  $\mbf{H}_A= \mbf{P}_a\otimes (\mbf{1}_b\trans/b)$ leading to $\mathcal H_{0}^{A} : \{\mbf{H}_A\mbf{C} = \mbf{0}\} = \{\bar{C}_{1\cdot}=\dots=\bar{C}_{a\cdot}\}$ 
	
	\item \textit{No main effect B}: $\mbf{H}_B= (\mbf{1}_a\trans/a)\otimes\mbf{P}_b$ leading to $\mathcal H_{0}^{B} : \{\mbf{H}_B\mbf{C} = \mbf{0}\} = \{\bar{C}_{\cdot 1}=\dots=\bar{C}_{\cdot b}\}$ 
	
	\item \textit{No interaction effect}:  $\mbf{H}_{AB}= \mbf{P}_a\otimes \mbf{P}_b$ and $\mathcal H_{0}^{AB} : \{\mbf{H}_{AB}\mbf{C} = \mbf{0}\} = \{\bar{C}_{\cdot \cdot} - \bar{C}_{\cdot i_B} - \bar{C}_{i_A\cdot} + \bar{C}_{i_Ai_B} = 0\}$  
\end{itemize}
Here, $\bar{C}_{i_A\cdot}=b^{-1}\sum_{i_B=1}^bC_{i_Ai_B}$ is the mean over the dotted index, and $\bar{C}_{\cdot i_B}$, $\bar{C}_{\cdot\cdot}$ are defined in the same way. A more lucid way to describe the aforementioned null hypotheses is based on an additive effect notation. Therefore, the MCV $C_{i_Ai_B} = C_0 + C_{i_A}^{\alpha} + C_{i_A}^{\beta} + C_{i_Ai_B}^{\beta}$ is decomposed into a general effect $C_0$, main effects $C_{i_A}^\alpha$ and $C_{i_B}^\beta$ of the factors $A$ and $B$, respectively, and the interaction effect $C_{i_Ai_B}^{\alpha\beta}$ under the usual side conditions $\sum_{i_A}C_{i_A}^\alpha = \sum_{i_B}C_{i_B}^{\beta} = \sum_{i_A}C_{i_Ai_B}^{\alpha\beta} = \sum_{i_B}C_{i_Ai_B}^{\alpha\beta} = 0$. Having this at hand, the null hypotheses can be rewritten as 
$\mathcal H_{0}^{A}: \{\mbf{H}_A\mbf{C} = \mbf{0}\} = \{C_{i_A}^\alpha = 0$ for all $i_A\}$ or $\mathcal H_{0,C}^{AB}: \{\mbf{H}_{AB}\mbf{C} = \mbf{0}\} = \{C_{i_Ai_B}^{\alpha\beta} = 0$ for all $i_A,i_B \}$.

\textit{Simulation settings}: We consider a $2\times 4$-design, i.e. $a=2$ and $b=4$ leading to $k=8$ different subgroups.  For the MVCs $C_{i_Ai_B}$, we choose three different scenarios. Under the first and second scenario, just a main effect of factor $A$ and $B$, respectively, is present, e.g., the null hypothesis $\mathcal H_0^A$ or $\mathcal H_0^B$, respectively, is false while the remaining two out of the three null hypotheses are true. In the third scenario, none of the three null hypotheses is true. For each scenario, we differentiate between two parameter constellations $A_1$ and $A_2$ resulting in six different settings, these are displayed in Table \ref{table1}. The simulation results are presented in Table \ref{table6}. We want to point out that the null hypotheses based on $C$ and $B$ are, in general, not equivalent as in the one-way layout. However, in the six settings considered here the null hypotheses for $B$ and $C$ are true at the same time or false at the same time. That is why we write, for ease of presentation, just $\mathcal H_0^A$ instead of $\mathcal H_{0,C}^A$ and $\mathcal H_{0,B}^B$.

\begin{table}
	\centering
	\caption{Six different settings for the choice of $C_{i_Ai_B}$ in a univariate $2\times 4$-layout}\label{table1}\scriptsize
	\scriptsize
	\begin{tabular}{ccc|cccc|cccc|cccc|}
		\hline
		& & &\multicolumn{4}{l}{$\mathcal H_0^B, \mathcal H_0^{AB}$ are true ($\mathcal H_0^{A}$ is false)} &\multicolumn{4}{l}{$\mathcal H_0^A,\mathcal H_0^{AB}$ are true ($\mathcal H_0^{B}$ is false)} &\multicolumn{4}{l}{--- (all hypotheses are false)}\\ 
		&$i_A$&$i_B$&$1$&$2$&$3$&$4$&$1$&$2$&$3$&$4$& $1$&$2$&$3$&$4$\\
		$A_1$&$1$&&0.237&0.237&0.237&0.237&0.237&0.300&0.300&0.300 &0.167&0.249&0.237&0.249\\
		&$2$&&0.300&0.300&0.300&0.300&0.237&0.300&0.300&0.300 &0.277&0.300&0.320&0.300\\
		\hline
		&$i_A$&$i_B$&$1$&$2$&$3$&$4$&$1$&$2$&$3$&$4$&$1$&$2$&$3$&$4$\\
		$A_2$&$1$&&0.346&0.346&0.346&0.346&0.346&0.300&0.300&0.300 &0.500&0.373&0.346&0.373\\
		&$2$&&0.300&0.300&0.300&0.300&0.346&0.300&0.300&0.300 &0.305&0.300&0.320&0.300\\
		\hline
	\end{tabular}
\end{table}

\begin{table}
	\centering
	\caption{Empirical sizes and powers (in $\%$) of the new asymptotic ($\text{Asy}$) and permutation ($\text{Per}$) tests under univariate two-way layout}\label{table6}\scriptsize
	\scriptsize
	\begin{tabular}{ccc|cc|cc|cc|cc|cc|cc|}
		\hline
		&&&\multicolumn{4}{l}{$\mathcal H_0^B, \mathcal H_0^{AB}$ are true ($\mathcal H_0^A$ is false)} &\multicolumn{4}{l}{$\mathcal H_0^A,\mathcal H_0^{AB}$ are true ($\mathcal H_0^B$ is false)} &\multicolumn{4}{l}{--- (all hypotheses are false)}\\ 
		&&&\multicolumn{2}{c}{$A_1$}&\multicolumn{2}{c}{$A_2$} &\multicolumn{2}{c}{$A_1$}&\multicolumn{2}{c}{$A_2$} &\multicolumn{2}{c}{$A_1$}&\multicolumn{2}{c}{$A_2$}\\
		Distr&Hyp&Test&$C$&$B$&$C$&$B$ &$C$&$B$&$C$&$B$ &$C$&$B$&$C$&$B$\\
		\hline
		$\mathrm{PE}_2$&$\mathcal H_0^A$&$\text{Asy}$&\textbf{93.8}&93.5&49.2&48.3 &{\bf 6.7}&6.3&5.1&5.3 &\textbf{97.6}&99.1&92.7&88.4\\
		&&$\text{Per}$&93.0&92.4&46.9&44.8 &5.8&5.0&4.5&5.0 &97.4&98.6&92.4&86.9\\
		&$\mathcal H_0^B$&$\text{Asy}$&{\bf 7.2}&{\bf 7.3}&6.1&6.0 &\textbf{73.3}&\textbf{67.8}&27.9&29.4 &\textbf{66.9}&\textbf{80.5}&30.4&23.9\\
		&&$\text{Per}$&5.2&5.5&4.7&4.9 &68.5&63.1&23.4&26.1 &58.1&74.0&24.1&18.3\\
		&$\mathcal H_0^{AB}$&$\text{Asy}$&{\bf 7.6}&6.2&{\bf 6.8}&6.3 &6.3&5.9&6.2&5.1 &\textbf{25.6}&48.9&\textbf{36.5}&31.0\\
		&&$\text{Per}$&5.9&4.7&5.4&5.3 &5.2&5.3&4.8&4.4 &20.4&41.7&32.5&24.5\\
		\hline
		$N$&$\mathcal H_0^A$&$\text{Asy}$&\textbf{81.6}&\textbf{81.6}&37.3&\textbf{38.1}&{\bf 6.6}&{\bf 7.2}&5.8&{\bf 7.4}&\textbf{93.5}&\textbf{95.8}&81.5&\textbf{76.9}\\
		&&$\text{Per}$&79.1&77.5&34.3&33.3 &5.6&5.4&4.9&4.8 &92.1&94.1&79.0&73.0\\
		&$\mathcal H_0^B$&$\text{Asy}$&{\bf 6.5}&{\bf 7.7}&{\bf 7.4}&{\bf 7.9} &\textbf{55.7}&\textbf{52.2}&\textbf{21.9}&\textbf{24.3} &\textbf{52.3}&\textbf{62.0}&\textbf{26.7}&\textbf{23.6}\\
		&&$\text{Per}$&4.9&4.6&4.9&4.7 &49.0&42.2&15.8&18.6 &44.5&51.8&19.3&15.1\\
		&$\mathcal H_0^{AB}$&$\text{Asy}$&6.2&{\bf 6.6}&{\bf 7.8}&{\bf 8.7} &{\bf 7.2}&{\bf 7.5}&{\bf 8.2}&{\bf 8.7} &\textbf{18.2}&\textbf{36.7}&\textbf{29.5}&\textbf{27.1}\\
		&&$\text{Per}$&4.8&4.6&6.4&6.0 &5.1&4.7&5.8&6.0 &12.9&28.6&25.2&18.9\\
		\hline
		$\mathrm{PE}_{.5}$&$\mathcal H_0^A$&$\text{Asy}$&\textbf{56.2}&\textbf{60.5}&\textbf{20.3}&\textbf{24.0} &{\bf 6.9}&{\bf 10.0}&{\bf 7.4}&{\bf 11.0} &\textbf{69.9}&\textbf{79.3}&\textbf{54.1}&\textbf{53.6}\\
		&&$\text{Per}$&50.6&48.8&16.6&16.5 &4.2&5.0&5.4&5.8 &65.6&71.8&49.7&42.7\\
		&$\mathcal H_0^B$&$\text{Asy}$&{\bf 11.0}&{\bf 14.9}&{\bf 11.2}&{\bf 14.6} &\textbf{37.7}&\textbf{39.9}&\textbf{20.2}&\textbf{25.7} &\textbf{35.4}&\textbf{43.8}&\textbf{19.8}&\textbf{21.7}\\
		&&$\text{Per}$&5.3&6.3&5.6&5.1 &24.2&23.1&10.4&12.5 &26.1&32.1&12.4&12.0\\
		&$\mathcal H_0^{AB}$&$\text{Asy}$&{\bf 8.8}&{\bf 14.7}&{\bf 10.1}&{\bf 15.6} &{\bf 9.9}&{\bf 14.9}&{\bf 8.9}&{\bf 12.2} &\textbf{16.9}&\textbf{29.6}&\textbf{21.1}&\textbf{23.8}\\
		&&$\text{Per}$&5.5&5.9&5.5&4.9 &5.7&5.3&4.7&4.5 &12.5&19.1&15.5&13.7\\\hline
		$t_5$&$\mathcal H_0^A$&$\text{Asy}$&\textbf{58.2}&\textbf{64.0}&{22.4}&\textbf{27.6} &{\bf 7.8}&{\bf 11.0}&5.9&{\bf 10.1} &\textbf{71.3}&\textbf{80.8}&{54.8}&\textbf{55.9}\\
		&&$\text{Per}$&53.9&54.0&18.1&18.7 &5.6&5.9&{\bf 3.5}&4.4 &66.8&73.3&50.2&46.4\\
		&$\mathcal H_0^B$&$\text{Asy}$&{\bf 10.0}&{\bf 16.4}&{\bf 9.9}&{\bf 15.1} &\textbf{39.0}&\textbf{41.9}&\textbf{20.4}&\textbf{29.0} &\textbf{38.7}&\textbf{49.8}&\textbf{20.9}&\textbf{25.3}\\
		&&$\text{Per}$&5.3&6.4&4.9&4.6 &29.1&24.1&11.0&13.6 &29.7&35.7&14.4&12.2\\
		&$\mathcal H_0^{AB}$&$\text{Asy}$&{\bf 7.6}&{\bf 16.4}&{\bf 10.0}&{\bf 16.2} &{\bf 8.6}&{\bf 16.7}&{\bf 8.7}&{\bf 16.5} &\textbf{13.8}&\textbf{34.1}&\textbf{21.9}&\textbf{28.0}\\
		&&$\text{Per}$&4.6&6.4&5.7&5.8 &4.8&5.4&5.2&5.7 &9.1&20.9&16.5&14.7\\
		\hline
	\end{tabular}
	Here, $C$ and $B$ denote that the tests for CVs and for inverses of CVs are used respectively. The empirical sizes are displayed in bold, when they are outside the $95\%$ significance limits, i.e., $[3.6\%,6.4\%]$. The empirical powers are in bold for too liberal tests, i.e., when the corresponding empirical sizes (or one of them for testing for interaction effects) are greater then $6.4\%$
\end{table}

\textit{Simulation results}: The empirical size of both permutation tests lies in the $95\%$ confidence interval $[3.6\%,6.4\%]$ in all null settings with just one conservative exception of $3.5\%$. At the same time, the asymptotic tests exhibit an acceptable  type-1 error behavior with a tendency to liberal decisions just under the $\mathrm{PE}_2$-distribution. Switching to the normal distribution, the liberality becomes more pronounced with values up to $8.7\%$ and an average of $7.3\%$. Under the $\mathrm{PE}_{.5}$- and $t_5$-distribution the tendency to liberality becomes even more extreme with values up to $14.9\%$ and $16.7\%$, respectively. In summary, it may be said that a larger kurtosis leads to more liberal decisions of the asymptotic test. Taking this 'unfair advantage' of the asymptotic test into account, the permutation tests show a reasonable power behavior. The usual gap between the asymptotic and permutation test is around $5\%$ and increases for the settings, where the asymptotic test shows an extreme liberal behavior under the corresponding 'true null' scenario or the two 'true null' scenarios for the interaction effect. \\
\textbf{Recommendation:} As in the previous section, we can only recommend the permutation tests, which keep the type-1 error rate accurately and show a reasonable power behavior.

\section{Illustrative real data examples}
\label{sec:dataexample}

We illustrate the practical application of our tests in practice on two real data sets. 

\subsection{Parkinson's disease data set}
First, we consider the multivariate case using the Parkinson's disease data set available at the UCI Machine Learning Repository \citep{FrankAsuncion2010}. This data set was created by Max Little of the University of Oxford, in collaboration with the National Centre for Voice and Speech, Denver, Colorado, who recorded the speech signals \citep{LittleEtAl2007, LittleEtAl2009}. 

There are $n=195$ observations of 22 variables, which are biomedical voice measures, e.g., the first variable is the average vocal fundamental frequency. The observations correspond to voice recordings from patients, which are divided into $k=2$ groups. The first sample of size $n_1=147$ consists of data for patients with Parkinson's disease (PD), while the second sample contains $n_2=48$ observations of healthy individuals. Based on these data, we want to discriminate healthy people from those with PD. 

For illustrative purposes, it is interesting to check if the multivariate coefficient of variations for two groups of healthy and ill people are significantly different. To this end, we apply the new asymptotic and permutation tests as well as the tests of \cite{AertsHaesbroeck2017}, supposing that the normal distribution is underlying, to infer the null hypothesis $\mathcal H_{0,C}:C_1=C_2$ using separately the first $2,3$ and $4$ variables of the data set. The results for higher dimensions (i.e., $d\geq 5$) are similar as for $d=4$, so they are omitted. The values of estimators of MCVs (in $\%$) are as follows $(\widehat{C}_1,\widehat{C}_2)=(22.18,26.61)$, $(20.60,26.42)$, $(20.13,20.28)$ for the first $2,3$ and $4$ variables, respectively.

For variables 1-2 ($d=2$) and 1-3 ($d=3$), the MCVs seem to be significantly different, while for variables 1-4 ($d=4$) this is not the case. The $p$-values of all tests are presented in Table \ref{table7}. 

\begin{table}
	\centering
	\caption{P-values (in $\%$) of the new asymptotic ($\varphi_{n,C}$, $\varphi_{n,B}$) and permutation ($\varphi_{n,C}^\pi$, $\varphi_{n,B}^\pi$) tests as well as the tests of \cite{AertsHaesbroeck2017} for Parkinson's disease data set for the first 2 (1-2), 3 (1-3) and 4 (1-4) variables}\label{table7}
	\begin{tabular}{lllllllllllll}
		\hline\noalign{\smallskip}
		&$\varphi_{n,C}$&$\varphi_{n,B}$&$\varphi_{n,C}^\pi$&$\varphi_{n,B}^\pi$&$AH_{C,C}$&$AH_{C,B}$&$AH_{R,C}$&$AH_{R,B}$&$AH_{S,C}$&$AH_{S,B}$&$AH_{SP,C}$&$AH_{SP,B}$\\
		\noalign{\smallskip}\hline\noalign{\smallskip}
		2&6.1&3.8&5.0&2.4&16.6&12.8&29.4&25.1&17.6&13.3&41.7&37.5\\
		3&1.2&0.5&1.8&0.4&6.4&3.5&15.3&10.8&5.0&2.3&17.5&13.1\\
		4&94.4&94.4&95.1&95.1&95.1&95.1&94.7&94.7&98.3&98.3&96.7&96.7\\
		\noalign{\smallskip}\hline
	\end{tabular}
\end{table}

As expected, none of the tests detect a significant difference of the MCVs when the first $d=4$ variables are considered. Switching to $d=2$, just our tests based on $B$ reject the null hypothesis while the tests based on $C$ are slightly above the $5\%$-benchmark but would lead at least for the level $\alpha=10\%$ to rejections. In contrast to the latter, the $p$-values of the competing tests of \cite{AertsHaesbroeck2017} range from $12.8\%$ up to $37.5\%$. In the remaining case $d=3$, the gap between the two MCV estimates becomes even wider explaining the overall smaller $p$-values. While the $p$-values for the proposed methods are below $2\%$, the decisions of the AH tests are diverse; only the $AH_{C,B}$, $AH_{S,C}$ and $AH_{S,B}$ tests reject the null hypothesis. The diversity in decisions of our and the $AH$ methods may be explained by misspecification of the underlying scenario for the $AH$ tests; the assumption of normality is maybe heavily violated. We checked the normality assumption graphically (plots not shown) and by applying the Shapiro-Wilk test for all four variables separately. Both support our aforementioned suspicion; the largest $p$-values of the Shapiro-Wilk tests was $1.8\cdot 10^{-8}$. Maybe, a certain transformation, e.g. using the logarithm, can fix this problem and yield to a better fit of the data to the normal distribution. Nevertheless, this example illustrates the benefit of having a nonparametric approach: the underlying distribution does not need to be known in advance.

\subsection{Beat the Blues data}
Let us now consider a univariate two-way layout. For this purpose, we re-analyse the BtheB data set \citep{ProudfootEtAl2003} from the R package HSAUR \citep{EverittHothorn2017}. The data were obtained in a clinical trial of an interactive multimedia program called ``Beat the Blues''. This program was designed to deliver cognitive behavioral therapy to $n=100$ depressed patients via a computer terminal. For illustrative purposes, we restrict to the following three variables, which were observed for all patients: drug - the patient takes anti-depressant drugs or not (factor with two levels: No, Yes); length - the length of the current episode of depression (factor with two levels: ${<6m}$ - less than six months, ${>6m}$ - more than six months); bdi.pre - Beck Depression Inventory II before treatment (quantitative variable). Using the proposed methods, we want to check whether taking drugs (factor A) and/or the duration of the current depression (factor B) have a significant effect on the patient's depression severity, which is measured by the variable bdi.pre. As explained in Section \ref{sec:setup}, the two-way layout can be incorporated by splitting the sample into $k=4$ subgroups with sample sizes $n_{11}=24$, $n_{12}=32$, $n_{21}=25$ and $n_{22}=19$, where the subgroup index $(i_A,i_B)$ is coded as follows: $i_A=1$ for drug-taking patients and $i_B=1$ for duration ${<6m}$. The $p$-values of the proposed tests for the two main effects as well as for the interaction effect are displayed in Table \ref{table8}.
\begin{table}
	\centering
	\caption{P-values (in $\%$) of the asymptotic ($\varphi_{n,C}$, $\varphi_{n,B}$) and permutation ($\varphi_{n,C}^\pi$, $\varphi_{n,B}^\pi$) tests for ``Beat the Blues'' data}\label{table8}
	\begin{tabular}{lllll}
		\hline\noalign{\smallskip}
		&$\varphi_{n,C}$&$\varphi_{n,B}$&$\varphi_{n,C}^\pi$&$\varphi_{n,B}^\pi$\\
		\noalign{\smallskip}\hline\noalign{\smallskip}
		drug&38.1&79.9&38.9&79.3\\
		length&1.4&1.7&1.1&1.8\\
		drug:length&2.5&3.3&3.2&4.0\\
		\noalign{\smallskip}\hline
	\end{tabular}
\end{table}
All asymptotic and permutation tests found an significant main effect of the variable length. However, the additionally observed significant interaction effect makes this conclusion uncertain. The concrete estimates of the MCVs and the standardized means displayed in Table \ref{table:C_exp2} indicate that the length variable has indeed no systematic effect; an effect is just present for the patients without drug influence. 
\begin{table}
	\centering
	\caption{The estimated MCVs and their reciprocals in the ``Beat the Blues'' data set}\label{table:C_exp2}
	\begin{tabular}{l|ll|ll}
		\hline\noalign{\smallskip}
		\multicolumn{1}{c}{}& \multicolumn{2}{c}{$\widehat C$} &\multicolumn{2}{c}{$\widehat B$}\\ 
		& $<6m$&$>6m$    & $<6m$ & $>6m$  \\
		drugs &  $41.28$  & $40.67$  &  2.33 & 2.46   \\
		no drugs & $58.41$ & $33.44$ & 1.71 & 2.99  \\
		\noalign{\smallskip}\hline
	\end{tabular}
\end{table}

\section{Conclusions}
\label{sec:conclusion}
We proposed generally applicable inference methods for the multivariate coefficient of variation (MCV) and its reciprocal, the standardized mean, in the general framework of potentially heteroscedastic factorial designs. Thus, not only one-way layouts but also higher-way layouts are covered allowing the discussion of main and interaction effects. While neither the multivariate nor the univariate coefficient of variation was considered in higher-way layouts, the current competitors, suggested by \cite{AertsHaesbroeck2017}, in the multivariate one-way layout set-up rely on restrictive assumptions concerning the underlying distribution. The advantage of our new tests is that no prior knowledge of the underlying situation is needed. The price of their broader applicability is the very unstable type-1 error control for small sample sizes. However, this can be solved by a permutation strategy. The resulting permutation methods are finitely exact under exchangeable data scenarios and still asymptotically valid for the general null hypotheses. Moreover, they are shown to be consistent under general alternatives.

In addition to these favorable theoretical findings, our extensive simulation study shows a significant benefit regarding type-1 error control of the permutation tests compared to the asymptotic approach and the competing methods of \cite{AertsHaesbroeck2017}. While the permutation tests can keep up with the others in terms of power in all settings, it clearly outperforms them in some cases, where the asymptotic approach and/or the existing methods of \cite{AertsHaesbroeck2017} lead to very conservative decisions under the respective null hypotheses. Consequently, we can only recommend the permutation tests, especially when no prior knowledge of the underlying situation is given and the sample sizes are small or moderate.

In this paper, we restricted our focus to the MCV definition of \cite{voinovNikulin1996}. This was mainly done because our competitors of \cite{AertsHaesbroeck2017} based on this definition as well and, thus, a fair comparison was possible. For our proofs, we first verified asymptotic normality of $(\mbf{\widehat \mu}_{i}, \mbf{\widehat\Sigma}_i)$ (resp. $(\mbf{\widehat \mu}_{1}^\pi, \mbf{\widehat\Sigma}_1^\pi,\ldots,\mbf{\widehat \mu}_{k}^\pi, \mbf{\widehat\Sigma}_k^\pi)$ for the permutation approach) and then applied a certain delta-method. The same strategy, just applying the delta-method for other functions, can be used to derive similar testing procedures for the other MCV definitions in \eqref{eqn:MCV}. Another possible aspect, we like to consider in the near future, is the robustification of the proposed permutation methods. For this purpose, we will follow \cite{AlbertZhang2010,AertsHaesbroeck2017,aertsETAL2018ellitical} and replace the empirical estimators for the mean and the covariance matrix by more robust estimators. 

\appendix

\section{Proofs}
\subsection{Proof of Theorem \ref{theo:conv_C+beta}}
For the proofs, we primarily apply the empirical process approach of \cite{vaartWellner1996} combined with the (functional) $\delta$-method. For the reader's convenience, we explain the technique briefly and refer to  \cite{vaartWellner1996} for an exhaustive introduction.  Let $P_i$ be the distribution of $\mbf{X}_{i1}$ and $\epsilon_{\mbf{x}}$ be the Dirac measure centred at $\mbf{x}\in\R^d$, i.e., $\epsilon_{\mbf{x}}(A)=\I\{\mbf{x}\in A\}$. Moreover, introduce the (group-wise) empirical process ${\mathbb{P}_{ni}}= n_i^{-1}\sum_{j=1}^{n_i}\epsilon_{\mbf{X}_{ij}}$. Define  $f_r(\mbf{x}) = x_r$ and $g_{rs}(\mbf{x})=x_rx_s$ for $r,s=1,\ldots,d$ and $\mbf{x}=(x_1,\ldots,x_d)\trans \in\R^d$. Subsequently, the measures $P_i$ and $\mathbb P_{ni}$ are indexed by the function class $\mathcal{F} = \{f_1,\ldots,f_d,g_{11},g_{12},\ldots,g_{dd}\}$. In detail, $P_i$ and $\mathbb P_{ni}$ are identified by $\{\int f \,\mathrm{ d }P_i:f\in\mathcal F\}$ and $\{\int f \,\mathrm{ d }\mathbb P_{ni}:f\in\mathcal F\} = \{n_i^{-1}\sum_{j=1}^{n_i} f(\mbf{X}_{ij} ):f\in\mathcal F\}$. For abbreviation, we write $Pf$ instead of $\int f \,\mathrm{ d } P$ and analogously for $\mathbb P_{ni}$. From now on we can treat $P_i$ and $\mathbb P_{ni}$ as random elements on $l^\infty(\mathcal F) = \{ Q\in\mathcal{M}_1(\R^d): \sup\{|Qf|:f\in\mathcal F\}<\infty\}$, where the space $\mathcal{M}_1(\R^d)$ consists of all probability measures on $\mathbb{R}^d$. Since $\mathcal F$ is a finite set, it is clearly a Vapnik-\u{C}ervonenkis-class, in short VC-class, and, hence, a Donsker class \citep[Sec. 2.6.1 and 2.6.2]{vaartWellner1996}. The latter implies
\begin{align}\label{eqn:donsker}
n_i^{1/2}( \mathbb{P}_{ni} - P_i) \overset{\mathrm d}{\longrightarrow} \mathbb{Z}_i \text{ on }l^\infty(\mathcal F),
\end{align}
where $\mathbb{Z}_i$ is a $P_i$-Brownian bridge. In particular,
\begin{align}\label{eqn:conv_dist_Pifr...}
&n_i^{1/2}\Big( n_i^{-1}\sum_{j=1}^{n_i} X_{ij1} - \mu_{i1}, \ldots,  n_i^{-1}\sum_{j=1}^{n_i} X_{ijd} - \mu_{id},\nonumber \\
&\phantom{n_i^{1/2}\Big(} n_i^{-1}\sum_{j=1}^{n_i} X_{ij1}X_{ij1} - \E(X_{i11}X_{i11}), \ldots,n_i^{-1}\sum_{j=1}^{n_i} X_{ij1}X_{ijd} - \E(X_{i11}X_{i1d}), \nonumber \\
&\phantom{n_i^{1/2}\Big(}\vdots \nonumber \\
&\phantom{n_i^{1/2}\Big(} n_i^{-1}\sum_{j=1}^{n_i} X_{ijd}X_{ij1} - \E(X_{i1d}X_{i1}), \ldots,n_i^{-1}\sum_{j=1}^{n_i} X_{ijd}X_{ijd} - \E(X_{i1d}X_{i1d}) \Big)\trans  \nonumber \\
&= n_i^{1/2} \Big( \mathbb{P}_{ni}f_1 - P_if_1,\ldots,\mathbb{P}_{ni}f_d - P_if_d, \mathbb{P}_{ni}g_{11} - P_ig_{11}, \mathbb{P}_{ni}g_{12} - P_ig_{12},\ldots, \mathbb{P}_{ni}g_{dd} - P_ig_{dd}\Big)\trans \nonumber \\
&\overset{\mathrm d}{\longrightarrow} \mbf{G}_i,
\end{align}
where $\mbf{G}_i=(G_{i1},\ldots,G_{i d'})\trans $ is centred, $d'$-dimensional normal distributed, $d'=d(d+1)$, with covariance structure
\begin{align*}
&\E(G_{ir}G_{is}) = \int f_rf_s \,\mathrm{ d }P_i - \int f_s \,\mathrm{ d }P_i \int f_r \,\mathrm{ d }P_i  = \E(X_{i1r}X_{i1s}) - \E(X_{i1r})\E(X_{i1s}) = [\mbf{\Sigma}_i]_{r,s},\\
&\E(G_{i(ad+r)}G_{is}) = \int g_{ar}f_s \,\mathrm{ d }P_i - \int g_{ar} \,\mathrm{ d }P_i \int f_s \,\mathrm{ d }P_i= \E(X_{i1a}X_{i1r}X_{i1s}) - \E(X_{i1a}X_{i1r})\E(X_{i1s}) = [\mbf{\Psi}_{i3}]_{ad-d+r,s}\\
&\E(G_{i(ad+r)}G_{i(bd+s)})  = \E(X_{i1a}X_{i1r}X_{i1b}X_{i1s}) - \E(X_{i1a}X_{i1r})\E(X_{i1b}X_{i1s}) = [\mbf{\Psi}_{i4}]_{ad-d+r,bd-d+s} 
\end{align*}
for $a,b,r,s\in\{1,\ldots,d\}$. In short,
\begin{align*}
\mbf{G}_i \sim N\Bigl( \mbf{0}, \begin{pmatrix}
\mbf{ \Sigma}_i & \mbf{\Psi}_{i3}\trans  \\
\mbf{\Psi}_{i3} & \mbf{\Psi}_{i4}
\end{pmatrix} \Bigr).
\end{align*}
To simplify the notation, we replace the prefactor $(n_i-1)^{-1}$ by $n_i^{-1}$ in the definition of the empirical covariance matrix estimator $\mbf{\widehat\Sigma}_{i}$. Clearly, this does not affect the asymptotic results. Now, define $\psi:\R^{d'}\to \R^{d'}$ by 
\begin{align*}
(\mbf{x}\trans ,\mbf{y}\trans )\trans =(x_1,\ldots,x_d, y_{11},y_{21},\ldots,y_{dd})\trans  \mapsto (x_1,\ldots,x_d, y_{11} - x_1 x_1, y_{21} - x_2x_1, \ldots, y_{dd} - x_dx_d)\trans .
\end{align*}
It is easy to check that $\psi$ is differentiable at every point $(\mbf{x}\trans ,\mbf{y}\trans )\trans $ with Jacobi matrix $\mbf{D}_\psi(\mbf{x})$ just depending on the first $d$ arguments $\mbf{x}=(x_1,\ldots,x_d)\trans $ given by
\begin{align*}
&\mbf{D}_\psi(\mbf{x}) =  \begin{pmatrix} \mbf{I}_d & \mbf{0}_{d\times d^2} \\ \mbf{\widetilde D}(\mbf{x})  & \mbf{I}_{d^2} \end{pmatrix},
\end{align*}
where $\mbf{\widetilde D}$ is defined in \eqref{eqn:def_tilde_D}, $\mbf{I}_d$ is the $d\times d$-dimensional unity matrix and $\mbf{0}_{d\times d^2}$ is the $d\times d^2$-dimensional zero matrix. The function $\psi$ even fulfills the following stronger differentiability condition, which is required later for the proof of Theorem \ref{theo:perm}, 
\begin{align}\label{eqn:psi_local_diff}
\frac{1}{t_n} \Big[\psi\begin{pmatrix} \mbf{x}_n + t_n\mbf{\widetilde x}_n \\ \mbf{y}_n + t_n\mbf{\widetilde y}_n\end{pmatrix} - \psi\begin{pmatrix} \mbf{x}_n \\ \mbf{y}_n \end{pmatrix}\Big] \to \mbf{D}_\psi(\mbf{x}) \begin{pmatrix}
\mbf{\widetilde x} \\
\mbf{\widetilde y}
\end{pmatrix}
\end{align}
for $\mbf{x}_n\to \mbf{x}\in\R^d$, $\mbf{\widetilde x}_n\to \mbf{\widetilde x}\in\R^d$, $\mbf{y}_n\to \mbf{y}\in\R^{d^2}$, $\mbf{\widetilde y}_n\to \mbf{\widetilde y}\in\R^{d^2}$, $t_n\to 0$. Combining the differentiability of $\psi$, the multivariate $\delta$-method \cite[Proposition 6.2]{bilodeauBrenner1999} and \eqref{eqn:conv_dist_Pifr...} we obtain 
\begin{align*}
n_i^{1/2} \begin{pmatrix} \mbf{\widehat \mu}_i - \mbf{\mu}_i \\ \text{vec}( \mbf{\widehat \Sigma}_i) - \text{vec}( \mbf{\Sigma}_i) \end{pmatrix} 
\overset{\mathrm d}{\longrightarrow} \mbf{D}_\psi(\mbf{\mu}_i) \mbf{G}_i.
\end{align*}
Now, introduce the map $\Phi:\R^d\times \text{GL}_{\text{sym}}(\R^d)\to \R$ defined as
\begin{align*}
\Phi(\mbf{a}, \mbf{A}) = \mbf{a}\trans \mbf{A}^{-1} \mbf{a}.
\end{align*}
Here, $\text{GL}_{\text{sym}}(\R^d)$ denotes the space of all nonsingular and symmetric $d\times d$-dimensional matrices. 
In the following, we treat a $d\times d$-dimensional matrix $\mbf{A}=(A_{ij})_{i,j=1,\ldots,d}$ as an element of $\R^{d^2}$ by vectorization $\text{vec}(\mbf{A})=(A_{11},\ldots,A_{d1},A_{21},\ldots,A_{dd})\trans $. In this spirit we endow the space $\text{GL}_{\text{sym}}(\R^d)$ by the Euclidean norm. To apply the $\delta$-method and its permutation version required later, $\Phi$ need to be differentiable in the stronger sense, analogously to \eqref{eqn:psi_local_diff}. For the reader's convenience, we explain briefly how this can be proven. First, let us have a close look at the inverse operation $\mbf{A}\mapsto \mbf{A}^{-1}$.  Let $\mbf{A}_n \to \mbf{A}\in \text{GL}_{\text{sym}}(\R^d)$, $\mbf{B}_n\to \mbf{B}\in\R^{d\times d}$ and $t_n\to 0$. Then
\begin{align}\label{eqn:diff_Ainv}
&t_n^{-1}\Bigl( (\mbf{A}_n + t_n \mbf{B}_n )^{-1} - \mbf{A}^{-1}_n \Bigr) =
t_n^{-1}\Bigl( (\mbf{I}_d + t_n \mbf{B}_n\mbf{A}_n^{-1} )^{-1}\mbf{A}_n^{-1} - \mbf{A}_n^{-1} \Bigr) \nonumber \\
&= t_n^{-1}\Bigl( \sum_{k=0}^\infty (-1)^k t_n^k (\mbf{A}_n^{-1}\mbf{B}_n)^{k} - \mbf{I}_k \Bigr)\mbf{A}_n^{-1} = - \mbf{A}_n^{-1}\mbf{B}_n\mbf{A}_n^{-1} + t_n \sum_{k=2}^\infty (-t_n)^{k-2}(\mbf{A}_n^{-1}\mbf{B}_n)^{k}\mbf{A}_n^{-1}\nonumber \\
&\to - \mbf{A}^{-1}\mbf{B}\mbf{A}^{-1},
\end{align}
where the convergence follows immediately for the simple case $\mbf{A}_n=\mbf{A}$. From the latter we can conclude the differentiability and the continuity of the inverse operation. Haven the continuity at hand, the convergence in \eqref{eqn:diff_Ainv} can be deduced for general sequences $\mbf{A}_n\to \mbf{A}$. The following relationship between the Kronecker product $\otimes$ and the vectorization operation is well-known \citep{neudecker1968}:  
\begin{align*}
\text{vec}(\mbf{\widetilde A}\mbf{\widetilde B} \mbf{\widetilde C})=( \mbf{\widetilde C}\trans \otimes  \mbf{\widetilde A}) \text{vec}( \mbf{\widetilde B})
\end{align*}
for all matrices $\mbf{\widetilde A}, \mbf{\widetilde B}, \mbf{\widetilde C}$ with appropriate dimensions such that the matrix multiplications are well defined. Combining this with \eqref{eqn:diff_Ainv} and using the abbreviation $\mbf{\widetilde A}_n = \mbf{A}_n + t_n\mbf{B_n}$, we obtain the differentiability of $\Phi$, even in the stronger sense required later for the permutation statement:
\begin{align}\label{eqn:phi_local_diff}
& t_n^{-1} \Bigl( (\mbf{a}_n+t_n\mbf{b_n})\trans (\mbf{A}_n + t_n \mbf{B}_n )^{-1}(\mbf{a}_n+t_n\mbf{b_n}) - \mbf{a}_n\trans \mbf{A}_n^{-1}\mbf{a}_n \Bigr)\nonumber \\
&= t_n^{-1} \Big( \mbf{a_n}\trans  \Big[ (\mbf{A}_n + t_n \mbf{B}_n )^{-1} - \mbf{A}_n^{-1} \Big] \mbf{a}_n + t_n\mbf{a}_n\trans \mbf{\widetilde A}_n^{-1}\mbf{b}_n + t_n\mbf{b}_n\trans \mbf{\widetilde A}_n^{-1}\mbf{a}_n + t_n^2\mbf{b}_n\trans \mbf{\widetilde A}_n^{-1}\mbf{b}_n \Big) \nonumber \\
&\to - \mbf{a}\trans \mbf{A}^{-1}\mbf{B}\mbf{A}^{-1}\mbf{a} + 2\mbf{a}\mbf{A}^{-1}\mbf{b} = - \Bigl( (\mbf{a}\trans \mbf{A}^{-1}) \otimes (\mbf{a}\trans \mbf{A}^{-1}) \Bigr)\text{vec}(\mbf{B}) + 2\mbf{a}\mbf{A}^{-1}\mbf{b}\nonumber \\
&= 	\mbf{D}_\Phi(\mbf{a}, \mbf{A})\begin{pmatrix}
\mbf{b}\\
\text{vec}(\mbf{B})	\end{pmatrix}
\end{align}
for $\mbf{A}_n \to \mbf{A}\in \text{GL}_{\text{sym}}(\R^d)$, $\mbf{B}_n\to \mbf{B}\in\R^{d\times d}$, $\mbf{a}_n\to \mbf{a}\in\R^d$, $\mbf{b}_n\to\mbf{b}\in\R^d$ and $t_n\to 0$, where the Jacobi matrix $\mbf{D}_\Phi$, identifying $\mbf{A}$ again by $\text{vec}(\mbf{A})$, is given by
\begin{align*}
\mbf{D}_\Phi(\mbf{a}, \mbf{A}) = \begin{pmatrix}
2 \mbf{a}\trans \mbf{A}^{-1} & -(\mbf{a}\trans \mbf{A}^{-1}) \otimes (\mbf{a}\trans \mbf{A}^{-1})
\end{pmatrix} .
\end{align*}
Hence, the $\delta$-method implies
\begin{align*}
n_i^{1/2} \Big( \mbf{\widehat \mu}_i\trans \mbf{\widehat \Sigma}_i^{-1}\mbf{\widehat\mu}_i - \mbf{ \mu}_i\trans \mbf{ \Sigma}_i^{-1}\mbf{\mu}_i \Big) \to \mbf{D}_\Phi(\mbf{\mu}_i, \mbf{\Sigma}_i)\mbf{D}_\psi(\mbf{\mu}_{i}) \mbf{G}_i.
\end{align*}
To prove the postulated asymptotic normality of ${C}_i$ and ${{B}}_i$, we apply the $\delta$-method to the functions $\varphi_1,\varphi_2:(0,\infty) \to \R$ defined by $\varphi_1(x)=x^{-1/2}$ and $\varphi_2(x)=x^{1/2}$, respectively, leading to
\begin{align*}
&n_i^{1/2} \Big( \widehat C_i - C_i \Big) \overset{\mathrm d}{\longrightarrow} -\frac{1}{2}(\mbf{ \mu}_i\trans \mbf{\Sigma}_i^{-1}\mbf{\mu}_i)^{-3/2} \mbf{D}_\Phi(\mbf{\mu}_i, \mbf{\Sigma}_i)\mbf{D}_\psi(\mbf{\mu}_{i}) \mbf{G}_i,\\
&n_i^{1/2} \Big( \widehat {B}_i - {B}_i \Big) \overset{\mathrm d}{\longrightarrow} \frac{1}{2}(\mbf{ \mu}_i\trans \mbf{\Sigma}_i^{-1}\mbf{\mu}_i)^{-1/2} \mbf{D}_\Phi(\mbf{\mu}_i, \mbf{\Sigma}_i)\mbf{D}_\psi(\mbf{\mu}_{i}) \mbf{G}_i.
\end{align*}
Since $n_i/n\to \kappa_i$ and $\mbf{A}(\mbf{\mu}_i,\mbf{\Sigma}_i) = \mbf{D}_\Phi(\mbf{\mu}_i, \mbf{\Sigma}_i)\mbf{D}_\psi(\mbf{\mu}_i)$, the distribution of the two limits coincide indeed with the ones of $Z_{i,C}$ and $Z_{i,{B}}$, respectively.

\subsection{Proof of Theorem \ref{theo:WTS_alt}}
We just present the proof of Theorem \ref{theo:WTS_alt}\eqref{enu:theo:WTS_alt_C}. The verification of \eqref{enu:theo:WTS_alt_beta} follows by just interchanging the letters $C$ and ${B}$. As already explained in the main paper, it is sufficient to check that $(\mbf{T}\mbf{C})\trans (\mbf{T}\mbf{\Sigma}_C\mbf{T}\trans )^+\mbf{T}\mbf{C}>0$ is positive under $\mathcal H_{1,C}:\mbf{T}\mbf{{B}}\neq \mbf{0}$. We first recall some well-known properties of the Moore--Penrose inverse for a quadratic matrix $\mbf{A}$: $(\mbf{A}\trans )^+ = (\mbf{A}^+)\trans $, $(\mbf{A}\trans \mbf{A})^+=\mbf{A}^+(\mbf{A}\trans )^+$ and $\mbf{A}\mbf{A}^+\mbf{A} = \mbf{A}$ \citep{raoMitra1971}. If the alternative $\mathcal H_{1,C}:\mbf{T}\mbf{{B}}\neq \mbf{0}$ is true, we can deduce from the nonsingularity of $\mbf{\Sigma}_C^{1/2}=\text{diag}(\sigma_{1,C},\ldots,\sigma_{k,C})$ that $\mbf{C}=\mbf{\Sigma}_C^{1/2}\mbf{v}$ for some non-zero vector $\mbf{v}\in\R^d\setminus\{\mbf{0}\}$. Altogether, 
\begin{align*}
\mbf{0} \neq \mbf{T}\mbf{C} = \mbf{T}\mbf{\Sigma}_C^{1/2}\mbf{v} = \mbf{T}\mbf{\Sigma}_C^{1/2}  (\mbf{T}\mbf{ \Sigma}_C^{1/2})^+\mbf{T} \mbf{\Sigma}_C^{1/2}\mbf{v} 
= \mbf{T}\mbf{ \Sigma}_C^{1/2} \Bigl[ (\mbf{T}\mbf{ \Sigma}_C^{1/2})^+\mbf{T}\mbf{C} \Bigr].
\end{align*}
Thus, $(\mbf{T} \mbf{\Sigma}_C^{1/2})^+ \mbf{T} \mbf{C}\neq \mbf{0}$ follows implying  
\begin{align*}
( \mbf{T} \mbf{C})\trans  ( \mbf{T} \mbf{\Sigma}_C \mbf{T}\trans  )^+ \mbf{T} \mbf{C} = ( \mbf{T} \mbf{C})\trans  ( \mbf{\Sigma}_C^{1/2} \mbf{T}\trans  )^+ ( \mbf{T} \mbf{  \Sigma}_C^{1/2} )^+ \mbf{T} \mbf{C} = \Bigl[ (\mbf{T}\mbf{\Sigma}_C^{1/2})^+\mbf{T}\mbf{C} \Bigr]\trans \Bigl[ (\mbf{T}\mbf{\Sigma}_C^{1/2})^+\mbf{T}\mbf{C} \Bigr] >0.
\end{align*}

\subsection{Proof of Lemma \ref{lem:var_equal_zero}}
For abbreviation, let
\begin{align}\label{eqn:def_tildeXi}
\mbf{\widetilde X}_i = ( X_{i11}, \ldots, X_{i1d}, X_{i11}X_{i11}, \ldots,  X_{i11}X_{i1d}, X_{i12}X_{i11}, \ldots, X_{i1d}X_{i1d} )\trans .
\end{align}
Note that the covariance matrix of $\mbf{\widetilde X}_i$ is given by
\begin{align*}
\mbf{\Sigma}_{\mbf{\widetilde X}} = \begin{pmatrix}
\mbf{ \Sigma}_i & \mbf{\Psi}_{i3}\trans  \\
\mbf{\Psi}_{i3} & \mbf{\Psi}_{i4}
\end{pmatrix}.
\end{align*}
Thus, $\sigma^2_{C,i} = 0$ implies that the distribution of $\mbf{A}(\mbf{\mu}_i,\mbf{\Sigma}_i)\mbf{\widetilde X}_i$ is degenerated. In other words, we have
\begin{align}\label{eqn:var=0_mat}
\begin{pmatrix} 2 \mbf{\mu}_i\trans \mbf{\Sigma}_i^{-1} - [(\mbf{\mu}_i\trans \mbf{\Sigma}_i^{-1}) \otimes (\mbf{\mu}_i\trans \mbf{\Sigma}_i^{-1})]\mbf{\widetilde D}(\mbf{\mu}_i),  & -(\mbf{\mu}_i\trans \mbf{\Sigma}_i^{-1}) \otimes (\mbf{\mu}_i\trans \mbf{\Sigma}_i^{-1})  \end{pmatrix}  \mbf{\widetilde X}_i
= \widetilde c
\end{align}
with probability one for some constant $\widetilde c\in\R$. Define for $r,s\in\{1,\ldots,d\}$
\begin{align*}
&a_r = [2 \mbf{\mu}_i\trans \mbf{\Sigma}_i^{-1} - \{(\mbf{\mu}_i\trans \mbf{\Sigma}_i^{-1}) \otimes (\mbf{\mu}_i\trans \mbf{\Sigma}_i^{-1})\}\mbf{\widetilde D}(\mbf{\mu}_i)]_r \in \R, \\
& b_{rs} = [-(\mbf{\mu}_i\trans \mbf{\Sigma}_i^{-1}) \otimes (\mbf{\mu}_i\trans \mbf{\Sigma}_i^{-1})]_{rd-d+s} =  -[\mbf{\mu}_i\trans \mbf{\Sigma}_i^{-1}]_s \cdot [\mbf{\mu}_i\trans \mbf{\Sigma}_i^{-1}]_r\in\R.
\end{align*}
Now, we can simplify \eqref{eqn:var=0_mat} to
\begin{align}\label{eqn:var=0_poly}
\sum_{r=1}^d( a_r X_{i1r} + b_{rr}X_{i1r}^2) + \sum_{s,r=1;s\neq r}^d b_{rs}X_{i1r}X_{i1s} = \widetilde c 
\end{align}
with probability one. Since $\mbf{\Sigma}_i$ is nonsingular and $\mbf{\mu}_i\neq \mbf{0}$, we have $\mbf{\mu}_i\trans \mbf{\Sigma}_i^{-1} \neq \mbf{0}$. Thus, $b_{rr}< 0$ holds for some $r\in \{1,\ldots,d\}$. Given the other components $(X_{i1s})_{s\neq r}$, the left hand side of \eqref{eqn:var=0_poly} is a polynomial in $X_{i1r}$ of degree two and, thus, $X_{i1r}$ can take at most two different values to solve \eqref{eqn:var=0_poly}.

\subsection{Proof of Theorem \ref{theo:perm}}
For the permutation sample, the groups are clearly not independent. That is why we need to discuss all groups together in a multivariate way. Let $P=\sum_{i=1}^{k}\kappa_iP_i$ be the pooled probability measure. Moreover, let $\mathbb{P}_n = n^{-1}\sum_{i=1}^{k}\sum_{j=1}^{n_i}\epsilon_{\mbf{X}_{ij}}=\sum_{i=1}^{k}(n_i/n)\mathbb{P}_{ni}$ be the empirical process of the pooled sample and $\mathbb{P}_{ni}^\pi = n_i^{-1}\sum_{j=1}^{n_i}\epsilon_{\mbf{X}_{ij}^\pi}$ be the permutation empirical process. Analogously to the proof of Theorem \ref{theo:conv_C+beta}, we index both processes by $\mathcal F$ and treat them as elements of $l^\infty(\mathcal F)$. \cite{vaartWellner1996} proved the permutation analogue of \eqref{eqn:donsker} for the case $k=2$, see their Theorems 3.7.1 and 3.7.2. The extension to $k\geq 3$ can be proven in a similar way \citep[Lemma 9 and Remark 1]{ditzhausETAL2019}. In our situation, we obtain from these results that given the observations almost surely we have the following conditional convergence in distribution:
\begin{align}\label{eqn:lem_NAE_perm_main_state}
n^{1/2}(\mathbb{P}_{n,1}^\pi-\mathbb{P}_n, \ldots,\mathbb{P}_{n,k}^\pi-\mathbb{P}_n) \overset{\mathrm d}{\longrightarrow} \mathbb{Z}^\pi\quad\text{on }(l^\infty(\mathcal F))^k,
\end{align}	
where $\mathbb{G}^\pi_{P}$ is a zero-mean Gaussian process on $(l^\infty(\mathcal F))^k$ with covariance function $\mbf{\Sigma}^\pi_Z: (l^\infty(\mathcal F))^k\times (l^\infty(\mathcal F))^k\to \R^{k\times k}$ defined for $\mbf{h}=(h_1,\ldots,h_k)\trans ,\mbf{\widetilde h}=(\widetilde h_1,\ldots,\widetilde h_k)\trans \in (l^\infty(\mathcal F))^k$ as
\begin{align}\label{eqn:cov_mat_Z}
&[\mbf{\Sigma}^\pi_Z(\mbf{h},\mbf{\widetilde h})]_{ii'}=  \gamma(i,i') \Bigl[ \int h_i\widetilde h_{i'}\,\mathrm{ d }P - \int h_i\,\mathrm{ d }P\int \widetilde h_{i'}\,\mathrm{ d }P \Bigr]
\quad\text{with}\quad \gamma(i,i')= \frac{1}{\kappa_i}\I\{i=i'\} - 1. 
\end{align}
From this we can obtain immediately the permutation analogue of \eqref{eqn:conv_dist_Pifr...}, where we just replace the original observations $\mbf{X}_{ij}$ by the permutation observations $\mbf{X}_{ij}^{\pi}$ and the expectations by their empirical pooled counterparts, e.g. $\E(X_{i1r})$ is replaced by $n^{-1}  \sum_{i,j}X_{ijr}$. To improve the readability, define
\begin{align*}
&\mathbb{G}_{ni}^\pi = \Big( \mathbb{P}_{ni}^\pi f_1 ,\ldots,\mathbb{P}_{ni}^\pi f_d, \mathbb{P}_{ni}^\pi g_{11} , \mathbb{P}_{ni}^\pi g_{12},\ldots, \mathbb{P}_{ni}^\pi g_{dd}\Big),\\
&\mathbb{G}_{n} = \Big(  \mathbb{P}_nf_1,\ldots, \mathbb{P}_nf_d,  \mathbb{P}_ng_{11},  \mathbb{P}_ng_{12},\ldots,  \mathbb{P}_ng_{dd}\Big).
\end{align*}
Consequently, we can deduce from \eqref{eqn:lem_NAE_perm_main_state} that given the observations almost surely
\begin{align}\label{eqn:Gni_pi}
n^{1/2} (\mathbb G_{n1}^\pi - \mathbb G_{n}, \ldots,\mathbb G_{nk}^\pi - \mathbb G_{n})\trans  \overset{\mathrm d}{\longrightarrow} \mbf{G}^\pi = ({\mbf{G}^\pi_1} \trans ,\ldots,{\mbf{G}^\pi_k }\trans )\trans ,
\end{align}
where $\mbf{G}^\pi$ is centred, $d^\pi$-dimensional normal distributed, $d^\pi= k d(d+1)$, and $\mbf{G}_i^\pi=(G_{11}^\pi,\ldots,G_{1d}^\pi)\trans $, $d'=d(d+1)$. Moreover, the covariance structure of  $\mbf{G}^\pi$ is given by
\begin{align*}
&\E(G_{ir}^\pi G_{i's}^\pi) = \gamma(i,i')\Bigl( \int f_rf_s \,\mathrm{ d }P - \int f_s \,\mathrm{ d }P \int f_r \,\mathrm{ d }P \Bigr)  = \gamma(i,i') \Bigl( \E(Y_rY_s) - \E(Y_r)\E(Y_s) \Bigr),\\
&\E(G_{i(ad+r)}^\pi G_{i's}^\pi) = \gamma(i,i')\Bigl( \int g_{ar}f_s \,\mathrm{ d }P - \int g_{ar} \,\mathrm{ d }P \int f_s \,\mathrm{ d }P  \Bigr) = \gamma(i,i')\Bigl( \E(Y_aY_rY_s) - \E(Y_aY_r)\E(Y_s) \Bigr) ,\\
&\E(G_{i(ad+r)}^\pi G_{i'(bd+s)}^\pi)  = \gamma(i,i')\Bigl( \E(Y_aY_rY_bY_s) - \E(Y_aY_r)\E(Y_bY_s) \Bigr), \quad a,b,r,s=1,\ldots,d;i,i'=1,\ldots,k,
\end{align*}
where $\mbf{Y}=(Y_1,\ldots,Y_d)\trans \in \R^d$ is a $P$-distributed random variable. In particular, we can see that each of the rescaled random vectors $\gamma(1,1)^{-1/2}\mbf{G}_1^\pi,\ldots,\gamma(k,k)^{-1/2}\mbf{G}_k^\pi$ has the same distribution, namely the centred $d'$-dimensional normal distribution with covariance matrix
\begin{align*}
\mbf{\Sigma}^\pi =  \begin{pmatrix}
\mbf{ \Sigma}_Y & \mbf{\Psi}_{Y3}\trans  \\
\mbf{\Psi}_{Y3} & \mbf{\Psi}_{Y4}
\end{pmatrix}.
\end{align*}
Here, $\mbf{\mu}_Y$ and $\mbf{\Sigma}_Y$ denote the expectation vector and the covariance matrix of $\mbf{Y}$, respectively. Moreover, the matrices $\mbf{\Psi}_{Y3}$ and $\mbf{\Psi}_{Y4}$ are defined as $\mbf{\Psi}_{i3}$ and $\mbf{\Psi}_{i4}$, respectively, while replacing $X_{i11},\ldots,X_{i1d}$ by $Y_1,\ldots,Y_d$. It is easy to check that $\mbf{\mu}_Y = \sum_{i=1}^k \kappa_i\mbf{\mu}_i = \mbf{\bar\mu} \neq \mbf{0}$ and 
\begin{lemma}\label{lem:sigmaY_pos+def}
	$\mbf{ \Sigma}_Y$ is positive definite.
\end{lemma}
A concrete proof of Lemma \ref{lem:sigmaY_pos+def} can be found below. To sum up, the covariance matrix of $\mbf{G}^\pi$ equals
\begin{align*}
\begin{pmatrix}
\gamma(1,1)\mbf{\Sigma}^\pi & \ldots & \gamma(1,k)\mbf{\Sigma}^\pi \\
\vdots & \ddots & \vdots \\
\gamma(k,1)\mbf{\Sigma}^\pi & \ldots & \gamma(k,k)\mbf{\Sigma}^\pi
\end{pmatrix} 
=\begin{pmatrix}
\kappa_1^{-1}\mbf{\Sigma}^\pi & \mbf{0}_{d'\times(k-2)d'} & \mbf{0}_{d'\times d'}\\[0.3em]
\mbf{0}_{(k-2)d'\times d'} & \ddots & \mbf{0}_{(k-2)d'\times d'} \\[0.3em]
\mbf{0}_{d'\times d'} & \ldots & \kappa_k^{-1}\mbf{\Sigma}^\pi
\end{pmatrix} 
- 
\begin{pmatrix}
\mbf{\Sigma}^\pi & \ldots & \mbf{\Sigma}^\pi\\
\vdots & \ddots & \vdots \\
\mbf{\Sigma}^\pi & \ldots & \mbf{\Sigma}^\pi
\end{pmatrix}.
\end{align*} 
To prove the asymptotic normality of $\mbf{\widehat C^\pi}$, we use again the $\delta$-method with the functions $\psi,\Phi,\varphi_1,\varphi_2$. Since centering is not done by fixed values, as in the proof of Theorem \ref{theo:conv_C+beta}, but by their empirical counterparts, which clearly depend on $n$, we need to apply a uniform version of the $\delta$-method \citep[Theorem 3.9.5]{vaartWellner1996}. For its application, the corresponding mappings need to be differentiable in the stronger sense of \eqref{eqn:psi_local_diff} and \eqref{eqn:phi_local_diff}, that we have already checked.  Now, let $\psi^{(k)}:\R^{kd'}\to \R^{kd'}$ be defined by $\psi^{(k)}(\mbf{z}_1,\ldots,\mbf{z}_k) = (\psi(\mbf{z}_1)\trans ,\ldots,\psi(\mbf{z}_k)\trans )\trans $ for $\mbf{z}_i\in\R^{d'}$. In the same way, we introduce $\Phi^{(k)},\varphi_1^{(k)},\varphi_2^{(k)}$. Moreover, let $\widehat C_0 = (\mbf{\widehat \mu}_0\trans \mbf{\widehat\Sigma}_0^{-1}\mbf{\widehat \mu}_0)^{-1/2}$ be the pooled counterpart of ${\widehat C}_i$, i.e. $\mbf{\widehat\mu}_0 = n^{-1}\sum_{i,j}\mbf{X}_{ij}$ and $\mbf{\widehat \Sigma}_0$ are the empirical expectation vector and the empirical covariance matrix of the pooled sample $\mbf{X}$, respectively. Note that $\mbf{\widehat\mu}_0\to \mbf{\mu}_Y$ and $\mbf{\widehat\Sigma}_0\to \mbf{\Sigma}_Y$, both almost surely. Lemma \ref{lem:sigmaY_pos+def} and $\mbf{\mu}_Y\neq \mbf{0}$ ensure that $\widehat C_0$ exists for sufficiently large $n$.  Combining the above arguments we obtain given the observations almost surely 
\begin{align*}
&n^{1/2} \Bigl( \widehat C_1^\pi - \widehat C_0,\ldots, \widehat C_k^\pi - \widehat C_0 \Bigr)\trans  = n^{1/2} \Bigl( \varphi_1^{(k)}(\Phi^{(k)}(\psi^{(k)}( \mathbb{G}_{n1}^\pi,\ldots,\mathbb{G}_{nk}^\pi ) )) - \varphi_1^{(k)}(\Phi^{(k)}(\psi^{(k)}(\mathbb{G}_{n},\ldots,\mathbb{G}_{n} ) )) \Bigr)\trans \\
& \overset{\mathrm d}{\longrightarrow} \begin{pmatrix}
\mbf{D}_C &
\mbf{0}_{1\times (k-2)d'} &	\mbf{0}_{1\times d'} \\[0.2em]
\mbf{0}_{(k-2)\times d'} & \ddots & 	\mbf{0}_{(k-2)\times d'}  \\[0.4em]
\mbf{0}_{1\times d'} & \mbf{0}_{1\times (k-2)d'} & \mbf{D}_C
\end{pmatrix}
\mbf{G}^\pi = \mbf{G}_C^\pi, \quad \mbf{D}_C = -\frac{1}{2}(\mbf{ \mu}_Y\trans \mbf{\Sigma}_Y^{-1}\mbf{\mu}_Y)^{-3/2} \mbf{D}_\Phi(\mbf{\mu}_Y, \mbf{\Sigma}_Y)\mbf{D}_\psi(\mbf{\mu}_Y).
\end{align*}
Clearly, $\mbf{G}_C^\pi$ is centred $k$-dimensional normal distributed with covariance matrix 
\begin{align*}
&\begin{pmatrix}
\mbf{D}_C &
\mbf{0}_{1\times (k-2)d'} &	\mbf{0}_{1\times d'} \\[0.2em]
\mbf{0}_{(k-2)\times d'} & \ddots & 	\mbf{0}_{(k-2)\times d'}  \\[0.4em]
\mbf{0}_{1\times d'} & \mbf{0}_{1\times (k-2)d'} & \mbf{D}_C
\end{pmatrix} 
\begin{pmatrix}
\gamma(1,1)\mbf{\Sigma}^\pi & \ldots & \gamma(1,k)\mbf{\Sigma}^\pi \\
\vdots & \ddots & \vdots \\
\gamma(k,1)\mbf{\Sigma}^\pi & \ldots & \gamma(k,k)\mbf{\Sigma}^\pi
\end{pmatrix}
\begin{pmatrix}
\mbf{D}_C\trans  &
\mbf{0}_{d'\times (k-2)} &	\mbf{0}_{d'\times 1} \\[0.2em]
\mbf{0}_{(k-2)d'\times 1} & \ddots & 	\mbf{0}_{(k-2)d'\times 1 }  \\[0.4em]
\mbf{0}_{d'\times 1} & \mbf{0}_{d'\times (k-2)} & \mbf{D}_C\trans 
\end{pmatrix} \\[0.5em]
&= \mbf{\widetilde \Sigma}_C - (\mbf{D}_C\mbf{\Sigma}^\pi\mbf{D}_C\trans ) \mbf{1}_{k\times k}, \quad \mbf{\widetilde \Sigma}_C = \text{diag}(\kappa_1^{-1}\mbf{D}_C\mbf{\Sigma}^\pi\mbf{D}_C\trans , \ldots,\kappa_k^{-1}\mbf{D}_C\mbf{\Sigma}^\pi\mbf{D}_C\trans ),
\end{align*}
where $\mbf{1}_{k\times k}$ is the $k\times k$-dimensional matrix consisting of $1$'s only. Since $\mbf{T}\mbf{1}_{k\times k}= \mbf{0}_{k\times k}$ we can deduce
\begin{align}\label{eqn:perm_C_conv}
n^{1/2} \mbf{T} \mbf{\widehat C^\pi} = n^{1/2} \mbf{T}( \mbf{\widehat C^\pi} - \widehat C_0 \mbf{1}_{k\times 1} ) \overset{ d}{\rightarrow} \mbf{G}_C^\pi\sim N( \mbf{0}, \mbf{T}\mbf{\widetilde \Sigma}_C\mbf{T})
\end{align}
given the observations almost surely. 
As needed for Theorem \ref{theo:WTS_null}, we need to ensure that $\mbf{\widetilde \Sigma}_C$ is nonsingular. Clearly, this is true if and only if $\mbf{D}_C\mbf{\Sigma}^\pi\mbf{D}_C\trans $ is not zero. The latter can be discussed in the same way as done in the proof of Lemma \ref{lem:var_equal_zero}.  It results that $\mbf{D}_C\mbf{\Sigma}^\pi\mbf{D}_C\trans = {0}$ implies that one component of $\mbf{Y}$ is conditionally two-point distributed. But it is easy to check that the latter is impossible under \assumption. To sum up, $\mbf{\widetilde\Sigma}_C$ is indeed nonsingular. Let us now consider the permutation covariance matrix estimator $\mbf{\widehat \Sigma}^\pi_C$. Since $\mathbb{P}_nf \to Pf$, $f\in \mathcal F$, almost surely according to the strong law of large numbers we can conclude from the continuous mapping theorem and \eqref{eqn:Gni_pi} that given the observations almost surely
\begin{align*}
\begin{pmatrix}
\mbf{\widehat \mu}_i^\pi  \\ \text{vec}( \mbf{\widehat \Sigma}_i^\pi) 
\end{pmatrix}
=
\psi(\mathbb{G}_{ni}^\pi)		  
\overset{p}{\longrightarrow}
\psi(\mathbb{P}_nf_1,\ldots, \mathbb{P}_ng_{dd}) =
\begin{pmatrix}
\mbf{\mu}_Y  \\ \text{vec}( \mbf{\Sigma}_Y)  \end{pmatrix} ,\qquad i=1,\ldots,k.
\end{align*}
In particular, we obtain that $\mbf{\widehat \Sigma}^\pi_{C}$ converges in probability to $\mbf{\widetilde \Sigma}_C$ and, thus, $(\mbf{T}\mbf{\widehat \Sigma}^\pi_{C}\mbf{T}\trans )^+$ converges in probability to $(\mbf{T}\mbf{\widetilde\Sigma}_{C}\mbf{T}\trans )^+$, both given the observations almost surely. Combining this, \eqref{eqn:perm_C_conv} and the continuous mapping theorem yields distributional convergence of $S_{n,C}^\pi(\mbf{T})$ to ${\mbf{G}_C^\pi }\trans (\mbf{T}\mbf{\widetilde\Sigma}_{C}\mbf{T}\trans )^+\mbf{G}^\pi$ given the observations almost surely, where the limit is, as postulated in Theorem \ref{theo:perm}, chi-squared distributed with $\text{rank}(\mbf{T})$ degrees of freedom \citep[Theorem 9.2.2]{raoMitra1971}. Repeating all the steps but replacing $\varphi_1$ by $\varphi_2$, we can deduce the analogue for $S_{n,{B}}^\pi(\mbf{T})$.

\section{Proof of Lemma \ref{lem:sigmaY_pos+def}}
Let $\mbf{z}=(z_1,\ldots,z_d)\trans \in \R^d\setminus\{\mbf{0}\}$. Recall that $\mbf{Y}$ is $P$-distributed with $P=\sum_{i=1}^d\kappa_iP_i$, where $P_i$ is the distribution of $\mbf{X}_{i1}$. In particular, we can deduce for any appropriate mapping $f$ that $\E(f(\mbf{Y})) = \int f(\mbf{y}) \,\mathrm{ d }P(\mbf{y}) = \sum_{i=1}^k \kappa_i\int f(\mbf{y}) \,\mathrm{ d }P_i(\mbf{y}) = \sum_{i=1}^k \kappa_i E(f(\mbf{X}_{i1}))$ holds. Moreover, note that $E\{[Z-E(Z)]^2\}\leq E\{[Z-a]^2\}$ for every $a\in \R$ and every real-valued random variable $Z$. Consequently,  
\begin{align*}
\mbf{z}\trans \mbf \Sigma_Y\mbf{z} &= \var\Big( \sum_{r=1}^dz_rY_r \Big) = \E\Big\{\Big[ \sum_{r=1}^dz_rY_r - \E\Big(\sum_{r=1}^dz_rY_r\Big)\Big]^2\Big\} \\
&= \sum_{i=1}^k \kappa_i \E\Big\{\Big[ \sum_{r=1}^dz_rX_{i1r} - \E\Big(\sum_{r=1}^dz_rY_r\Big)\Big]^2\Big\} \geq \sum_{i=1}^k \kappa_i \E\Big\{\Big[ \sum_{r=1}^dz_rX_{i1r} - \E\Big(\sum_{r=1}^dz_rX_{i1r}\Big)\Big]^2\Big\} \\
&= \sum_{i=1}^k \kappa_i \mbf{z}\trans \mbf \Sigma_i\mbf{z}> 0.
\end{align*}

\section*{Acknowledgement}
	Marc Ditzhaus was funded by the \textit{Deutsche Forschungsgemeinschaft} (grant no.  PA-2409 5-1). A part of calculations for simulation study was made at the Pozna\'n Supercomputing and Networking Center (grant no. 382).

\bibliographystyle{spbasic} 
\bibliography{sample}

\end{document}